\documentclass[twocolumn,showpacs,amsmath,amssymb,pra,superscriptaddress]{revtex4-1}
\usepackage{graphicx}
\usepackage{dcolumn}
\usepackage{bm}
\usepackage {longtable}
\usepackage{color}      
\usepackage{supertabular}

\expandafter\let\csname equation*\endcsname\relax
\expandafter\let\csname endequation*\endcsname\relax
\usepackage{amsmath}
\newcommand{\ket}[1]{|#1\rangle}
\newcommand \be{\begin{equation}}
\newcommand \ee{\end{equation}}
\newcommand \bea{\begin{eqnarray}}
\newcommand \eea{\end{eqnarray}}
\newcommand \bse{\begin{subequations}}
\newcommand \ese{\end{subequations}}
\begin{document}

\title {Simulated quantum process tomography of quantum gates with Rydberg superatoms}

\author{I.~I.~Beterov}
\email{beterov@isp.nsc.ru}
\affiliation {Rzhanov Institute of Semiconductor Physics SB RAS, 630090 Novosibirsk, Russia}
\affiliation {Novosibirsk State University, Interdisciplinary Quantum Center, 630090 Novosibirsk, Russia}
\affiliation {Novosibirsk State Technical University, 630073 Novosibirsk, Russia}

\author{M.~Saffman}
\affiliation {Department of Physics, University of Wisconsin, Madison, Wisconsin, 53706, USA}

\author{E.~A.~Yakshina}
\affiliation {Rzhanov Institute of Semiconductor Physics SB RAS, 630090 Novosibirsk, Russia}
\affiliation {Novosibirsk State University, Interdisciplinary Quantum Center, 630090 Novosibirsk, Russia}

\author{D.~B.~Tretyakov}
\affiliation {Rzhanov Institute of Semiconductor Physics SB RAS, 630090 Novosibirsk, Russia}
\affiliation {Novosibirsk State University, Interdisciplinary Quantum Center, 630090 Novosibirsk, Russia}

\author{V.~M.~Entin}
\affiliation {Rzhanov Institute of Semiconductor Physics SB RAS, 630090 Novosibirsk, Russia}
\affiliation {Novosibirsk State University, Interdisciplinary Quantum Center, 630090 Novosibirsk, Russia}

\author{G.~N.~Hamzina}
\affiliation {Rzhanov Institute of Semiconductor Physics SB RAS, 630090 Novosibirsk, Russia}
\affiliation {Novosibirsk State Technical University, 630073 Novosibirsk, Russia}

\author{I.~I.~Ryabtsev}
\affiliation {Rzhanov Institute of Semiconductor Physics SB RAS, 630090 Novosibirsk, Russia}
\affiliation {Novosibirsk State University, Interdisciplinary Quantum Center, 630090 Novosibirsk, Russia}

\begin{abstract}
We have numerically simulated quantum tomography of single-qubit and two-qubit quantum gates with qubits represented by mesoscopic ensembles containing random numbers of atoms.  Such ensembles of strongly interacting atoms in the regime of Rydberg blockade are known as Rydberg superatoms. The Stimulated Raman Adiabatic Passage (STIRAP) in the regime of Rydberg blockade is used for deterministic Rydberg excitation in the ensemble, required for storage of quantum information in the collective state of the atomic ensemble and implementation of two-qubit gates. The optimized shapes of the STIRAP pulses are used to achieve high fidelity of the population transfer. Our simulations confirm validity and high fidelity of single-qubit and two-qubit gates with Rydberg superatoms. 

\end{abstract}
\pacs{32.80.Ee, 03.67.Lx, 34.10.+x, 32.70.Jz , 32.80.Rm}
\maketitle

\section{Introduction}

Neutral atoms are promising candidates for building a quantum computer, since they meet all the DiVincenzo criteria for qubits~\cite{DiVincenzo2000}. A large array of optical dipole traps, loaded with single atoms, can be used as a scalable quantum register~\cite{Piotrowicz2013, Xia2015}. However, single-atom loading of the optical dipole traps remains technically challenging, and the unavoidable single-atom losses in the optical dipole traps will inevitably lead to computational errors. Another approach is based on storage of quantum information in the collective states of mesoscopic atomic ensembles or superatoms~\cite{Lukin2001}. Quantum information with Rydberg atoms commonly exploits the effect of Rydberg blockade, when only one atom in the ensemble of strongly interacting atoms can be excited into a Rydberg state by narrow-band laser excitation~\cite{Saffman2010,Lukin2001}. These ensembles, known as Rydberg superatoms~\cite{Lukin2001,Stanojevic2009, Heidemann2007}, can be considered as effective two-level systems with enhanced Rabi frequency  $\Omega =\Omega_0\sqrt N$, where  $\Omega_0$ is a single-atom Rabi frequency, and \textit{N} is the number of interacting atoms in the ensemble. The collective Rabi oscillations have been observed for two atoms~\cite{Urban2009,Wilk2010} and for large atomic ensembles~\cite{Dudin2012,Ebert2014,Zeiner2015}. One of the most important drawbacks of superatom qubits are the fluctuations of the number of atoms in the ensemble due to random loading of optical dipole traps.  This makes it difficult to implement high-fidelity quantum gates due to fluctuations of the collective Rabi frequency $\Omega$.

In our recent works we proposed to overcome this difficulty using adiabatic passage and Rydberg blockade for deterministic single-atom Rydberg excitation~\cite{Beterov2011} and the dynamic phase compensation~\cite{Beterov2013,Beterov2014}. Schemes of single-qubit and two-qubit gates for mesoscopic qubits have been proposed~\cite{Beterov2013,Beterov2014}. The aim of the present work is to confirm the validity of these schemes by numeric simulation of quantum tomography, and to estimate the maximum fidelity of the quantum gates which can be achieved with our approach using mesoscopic qubits. Quantum tomography is a powerful technique which is used for full reconstruction of the properties of quantum states and quantum processes using a sequence of specific measurements over qubits~\cite{Nielsen2011,James2001,Poyatos1997,White2007}. This technique has been successfully implemented in a number of experiments with trapped ions~\cite{Riebe2006,Roos2004}, superconducting qubits~\cite{Yamamoto2010}, nitrogen-vacancy qubits~\cite{Shukla2014}, NMR systems~\cite{Childs2001}, single photons~\cite{Mitchell2003}, etc. In this paper we have performed a numeric simulation of single-qubit and two-qubit state and process tomography with qubits represented by atomic ensembles containing \textit{N}=1-4 interacting atoms in the regime of Rydberg blockade.

This paper is organized as follows. In Section~2 we discuss the optimized schemes of quantum gates based on adiabatic passage and Rydberg blockade. Section~3 is devoted to numeric simulation of single-qubit and two-qubit quantum process tomography. In Section~4 the possible error sources are discussed. A review of single-qubit and two-qubit state and process tomography for two-level qubits is presented in the Appendix.

\begin{center}
\begin{figure}[!t]
 \center
\includegraphics[width=\columnwidth]{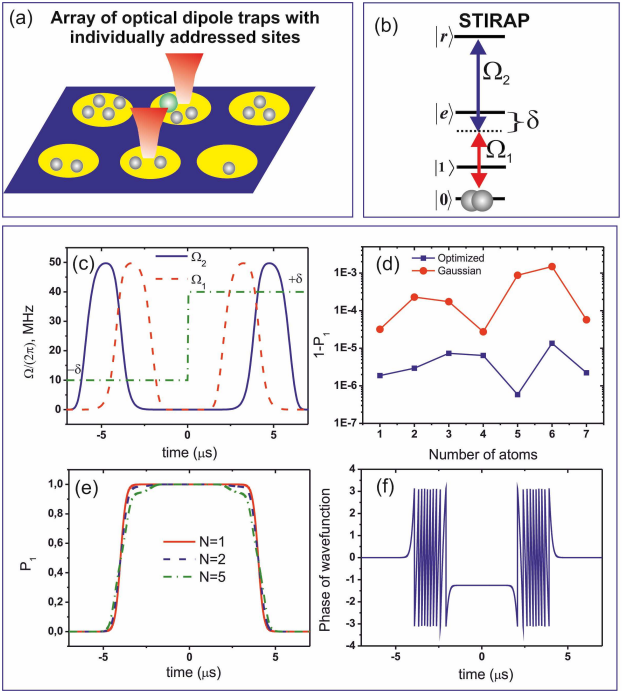}
\vspace{-.5cm}
\caption{
\label{Scheme}(Color online).
Scheme of a quantum register based on qubits represented by mesoscopic atomic ensembles; (b) Scheme of the typical energy levels for single-atom Rydberg excitation using STIRAP; (c) Shapes of the optimized STIRAP pulses~\cite{Vasiliev2009}; (d) Comparison of the error $1-P_1$ of population transfer for Gaussian and optimized STIRAP sequence; (e) Time dependence of the probability $P_1$ of single-atom Rydberg excitation during double STIRAP sequence in mesoscopic atomic ensembles for different number of atoms \textit{N}=1,2,5; (f) Phase of the probability amplitude of the ground state of mesoscopic ensemble.
}
\end{figure}
\end{center}

\section{Quantum gates based on optimized double adiabatic passage}

Our approach for building a quantum register is based on the array of randomly loaded optical dipole traps as shown in figure~\ref{Scheme}(a). We use a Stimulated Raman Adiabatic Passage (STIRAP) technique~\cite{Bergmann1998} for deterministic single-atom Rydberg excitation in a regime of a Rydberg blockade~\cite{Beterov2011}. This technique exploits counter-intuitive sequence of overlapping laser pulses in a three-level system at two-photon resonance, as shown in figure~\ref{Scheme}(b). Similar results can be obtained by using of single-photon adiabatic excitation with chirped laser pulses~\cite{Beterov2011, Kuznetsova2015, Liu2015, Malinovsky2001}.
Our scheme of quantum gates is based on a double adiabatic sequence, shown in figure~\ref{Scheme}(c), for laser excitation and subsequent de-excitation of the single Rydberg atom in the ensemble~\cite{Beterov2011,Beterov2013,Beterov2014}. High fidelity quantum gates require high fidelity of Rydberg excitation, but commonly used STIRAP techniques with Gaussian pulses usually provide the infidelity larger than 10\textsuperscript{-4} even in theory. The fidelity of the population transfer can be improved by optimization of the shapes of STIRAP pulses, as proposed in Ref.~\cite{Vasiliev2009}. 

We have used the following shapes of the optimized STIRAP pulses from~\cite{Vasiliev2009}:
\bea
\label{eq1}
\Omega_1\left(t\right)&=\Omega_0 F\left(t-t_1\right)\text{sin}\left[\frac{\pi} 2f\left(t-t_1\right)\right]&+\nonumber\\
&+\Omega_0 F\left(t-t_2\right)\text{cos}\left[\frac{\pi} 2f\left(t-t_2\right)\right]&  \nonumber \\
\Omega_2\left(t\right)&=\Omega_0 F\left(t-t_1\right)\text{cos}\left[\frac{\pi } 2f\left(t-t_1\right)\right]&+\nonumber\\
&+\Omega_0 F\left(t-t_2\right)\text{sin}\left[\frac{\pi } 2f\left(t-t_2\right)\right].&
\eea

\noindent Here  $F\left(t\right)=\text{exp}\left[-\left(t/T_0\right)^{2n}\right]$ and  $f\left(t\right)=\left[1+\text{exp}\left(-\mathit{\lambda t}/T\right)\right]^{-1}$ . Following Ref.~\cite{Vasiliev2009}, we have chosen  $T_0=2T$,  $n=3$, and  $\lambda =4$. In our calculations the Rabi frequency for both pulses is $\Omega _0/\left(2\pi \right)=50$ MHz, detuning from the intermediate state is $\delta /\left(2\pi \right)=200$ MHz, and $T_0=2\;\mathit{\mu s}$ is the time parameter for a hypergaussian function $F\left(t\right)$ which determines the pulse duration. The positions of the pulses are defined by $t_1=-4\,\mu s$ and $t_2=4\,\mu s$.

We have compared the fidelity of population inversion of the optimized STIRAP scheme with the conventional Gaussian pulses: 
\bea
\label{eq2}
\Omega _1\left(t\right)&=&\Omega _0\text{exp}\left[-\left(t-t_1\right)^2/2\tau ^2\right] \nonumber \\
\Omega _2\left(t\right)&=&\Omega _0\text{exp}\left[-\left(t-t_2\right)^2/2\tau ^2\right]. 
\eea
\noindent with  $t_1=1\;\mathit{\mu s}$,  $t_2=-1\;\mathit{\mu s}$ and  $\tau =1\;\mathit{\mu s}$.

Comparison of the numerically calculated fidelity of single-atom Rydberg excitation in the atomic ensemble consisting of \textit{N} atoms for Gaussian and optimized pulses is shown in figure~\ref{Scheme}(d). We have solved a Schr\"odinger equation for the probability amplitudes in a quasimolecule which consists of \textit{N} three-level atoms, interacting with two laser fields. The perfect Rydberg blockade was considered in the simulations by removing all quasimolecular states with more than one Rydberg excitation. The finite lifetimes of intermediate and Rydberg states have not been taken into account (this assumes short interaction times compared to lifetimes). The optimized pulse shapes allow substantial reduction of the infidelity of single-atom Rydberg excitation, which is kept below 10\textsuperscript{-5} for almost all cases, as shown in figure~\ref{Scheme}(d).

\begin{figure}[!t]
 \center
\includegraphics[width=\columnwidth]{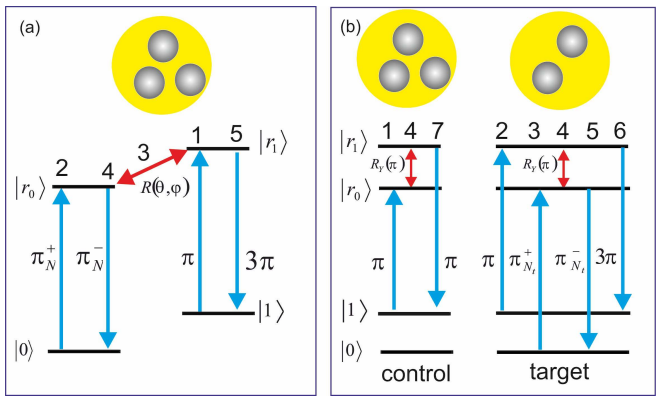}
\vspace{-.5cm}
\caption{
\label{Rotation}(Color online).
(a) Scheme of single-qubit rotation for a mesoscopic atomic ensemble with random number of atoms. Pulses 1-5 act between the qubit states $\ket{0}$, $\ket{1}$ and the Rydberg states $r_0$ and $r_1$. Pulses 2 and 4 are two-photon STIRAP sequences with opposite signs of the detuning from the intermediate state. Pulses~1 and 5 are coherent single-atom $\pi$ and $3\pi$ pulses. Pulse~3 is a microwave or Raman transition between Rydberg states $r_0$ and $r_1$ with arbitrary area $\theta$ and phase $\phi$. Only one Rydberg excitation in the ensemble is allowed due to Rydberg blockade. (b) Scheme of CNOT-type two-qubit gate with two mesoscopic atomic ensembles~\cite{Beterov2014}. Only one Rydberg atom can be excited in the whole system of interacting atoms due to Rydberg blockade. The pulses 2-6 invert the state of the target qubit if the control qubit is initially prepared in state $\ket{\bar{0}}$ and remains in its ground state during the whole pulse sequence. 
}
\end{figure}

The time dependences of the probability $P_1$ of single-atom Rydberg excitation, and of the phase of the probability amplitude of the ground state in the atomic ensemble interacting with two optimized STIRAP sequences are shown in figures~\ref{Scheme}(e) and \ref{Scheme}(f), respectively. The ensemble returns to the ground state after the end of the second STIRAP sequence, and the phase of the ground state wavefunction is preserved, but only in the case when the sign of the detuning from the intermediate excited state is switched between two STIRAP sequences \cite{Beterov2013,Beterov2014}, as shown in figure~\ref{Scheme}(c). The phase conservation allowed us to develop the schemes of high-fidelity single-qubit and two-qubit quantum gates with mesoscopic atomic ensembles \cite{Beterov2013,Beterov2014}, which are shown in figure~\ref{Rotation}.

The idea behind these ensemble gates is based on the following considerations: quantum information can be stored in the hyperfine sublevels of the ground state of alkali-metal atoms, denoted as  $\ket{0} $ and  $\ket{1} $. The ground state of a mesoscopic ensemble which consists of \textit{N} atoms is denoted as  $\ket{\bar {0}} =\frac{1}{\sqrt{N}}\ket{00...0} $. In the regime of Rydberg blockade we can use adiabatic passage to deterministically excite a collective state with a single Rydberg excitation  $\ket{\bar {1} } =\frac{1}{\sqrt{N}}\sum\limits_{j=1}^N\ket{00...r_j...0}$. This state can be then mapped onto the other hyperfine sublevel by a coherent single-atom  $\pi $ pulse. Therefore we consider the states $\ket{\bar 0} =\ket{00...0} $ and  $\ket{\bar 1} =\frac{1}{\sqrt{N}}\sum\limits_{j=1}^N \ket{00...1_j...0} $ as logical states of the ensemble qubit. We have to take into account the accumulation of the \textit{N}-dependent dynamic phase during the first adiabatic sequence, as shown in figure~\ref{Scheme}(f). If the adiabatic excitation of the ensemble into the Rydberg state is followed by coherent  $\pi $ pulse at   $\ket{r} \to \ket{1}$ transition, the final qubit state will be  $\ket{1}=\frac{1}{\sqrt{N}}e^{\mathit{i\alpha}_N}\sum\limits_{j=1}^N\ket{00...1_j...0}$ where  $\alpha_N$ is the accumulated phase, which depends on the number of atoms in the ensemble and parameters of laser excitation (Rabi frequencies and pulse shapes). Our simulations of quantum tomography (to be given below) confirm that this phase accumulation does not affect the gate performance, but this is true only in the case when switching the detuning from the intermediate level in the double STIRAP sequence is used.

A scheme of the single-qubit rotation around X and Y axes on a Bloch sphere is shown in figure~\ref{Rotation}(a). The states $r_0$ and  $r_1$ are two Rydberg levels. Strong Rydberg interaction ensures that in the ensemble there could be only one Rydberg excitation. Notably, excitation of two atoms into different Rydberg states  $r_0$ and  $r_1$ is also blocked due to strong Rydberg interaction.  $\pi _N^+$ and $\pi _N^-$ indicate STIRAP in the \textit{N}-atom ensemble with opposite signs of the detuning. The  $\pi $ and  $3\pi $ pulses are rotations  $R_X\left(\pi \right)$ and  $R_X\left(3\pi \right)$, respectively. The  $3\pi $ pulse is used instead of a  $\pi $ pulse for compensation of the additional phase shift of the state  $\ket{1}$, arising from combined action of two  $\pi $ pulses. The transition between two Rydberg levels $r_0$ and  $r_1$ is described by a Rabi rotation matrix:

\be
\label{eq3}
R\left(\theta ,\varphi \right)=\begin{pmatrix}
\text{cos}\frac{\theta}{2}&i\text{e}^{-i\varphi }\text{sin}\frac{\theta}{2}\\
i\text{e}^{i\varphi }\text{sin}\frac{\theta}{2}&\text{cos}\frac{\theta }{2}.
\end{pmatrix}.
\ee

\noindent The qubit rotations around X and Y axes on a Bloch sphere (des) can be represented as $R_x\left(\theta \right)=R\left(-\theta ,0\right)$ and  $R_y\left(\theta \right)=R\left(-\theta ,\pi /2\right)$. The scheme of a CNOT-type gate, shown in figure~\ref{Rotation}(b), is based on the effect of Rydberg blockade. It is a modification of the amplitude-swap gate for two atoms, which has been experimentally demonstrated in Ref.~\cite{Isenhower2010}. The pulse sequence~2-6 acting on a target qubit inverts its quantum state, but only in the case if the control qubit remains in state  $\ket{0} $ and is not excited into the Rydberg state. If the control qubit is initially prepared in the state  $\ket{1}$, its Rydberg excitation by pulse~1 blocks transitions to the Rydberg states for a target qubit, and leaves it in its initial state. The control qubit is returned back to the ground state by pulse~7. This scheme can be converted to a conventional CNOT by adding a NOT operation on the control qubit before and after the two-qubit gate.

\section{Simulated quantum tomography of single-qubit and two-qubit quantum gates}

In the present work we have performed a full numeric simulation of the single-qubit and two-qubit quantum process tomography for mesoscopic atomic ensembles. Initially all the ensembles are assumed to be prepared in the ground state  $\ket{\bar 0} =\ket{00...0} $. We have numerically simulated all procedure of quantum tomography including preparation of different basis states of the atomic ensembles, implementation of various quantum gates and tomographic measurements of the final quantum states. The basic principles of quantum tomography are described in Ref.~\cite{Nielsen2011} and the process fidelity of a Rydberg blockade gate between single atom 
qubits was simulated in Ref.~\cite{XZhang2012}. A brief review of quantum tomography for two-level qubits is presented in the Appendix.

\subsection{Quantum state tomography} 

The purpose of quantum state tomography is reconstruction of the density matrix of a two-level qubit. A single-qubit density matrix can be written as  $\rho_{\left(1\right)}=\frac{1}{2} \sum\limits_{i=1}^{4}\text{Tr}\left(\sigma_i\rho_{\left(1\right)}\right)\sigma_i$ where  
$\sigma_i$ are four Pauli matrices  $I=\sigma _0$,  $X=\sigma_x$, $Y=\sigma _y$ and  $Z=\sigma _z$. This means that we can express the quantum state through four quantum mechanical observables. Similar procedure is used for two-qubit state tomography. We represent two-qubit density matrix as  $\rho_{\left(2\right)}=\frac{1}{4}\sum\limits_{i,j=1}^{4}\text{Tr}\left[\left(\sigma_i \otimes \sigma_j\right)\rho_{\left(2\right)}\right]\left(\sigma_i\otimes\sigma _j\right)$ and perform 16 measurements of the observables.

\subsection{Quantum process tomography} 

The purpose of quantum process tomography is reconstruction of the quantum process, which transfers the initial state of a quantum system into the final state. Every quantum process can be considered as a transformation of the density matrix: 

\be
\label{eq4}
\rho'=\varepsilon \left(\rho \right).
\ee

\noindent For a fixed set of operators  $\tilde  E_i$ this expression can be rewritten as

\be
\label{eq5}
\rho' =\varepsilon \left(\rho \right)=\underset{i,j}{\sum }\chi_{ij}\tilde  E_i\rho \tilde  E_j^{\dag}.
\ee
\noindent Any quantum process can be represented by a  $\chi$-matrix which is  $4\times 4$ for single-qubit process and  $16\times 16$ for two-qubit operations. The quantum process tomography requires the following steps: (i) preparation of initial basis states of the quantum system; (ii) quantum operation with qubits prepared in all basis states; (iii) quantum state tomography of the final states of the qubits after the quantum process under study is finished.

\subsection{Maximum-likelihood reconstruction and fidelity}

Both quantum state tomography and quantum process tomography can lead to non-physical density or process matrices. A maximum-likelihood reconstruction~\cite{James2001,OBrien2004,Howard2006,Lvovsky2009} is the procedure which allows finding the correct matrix  $\tilde {\chi }$ which is closest to the measured one. To estimate the fidelity of the quantum state preparation or quantum gate we compare the reconstructed matrix with the ideal matrix  $\chi_{id}$ which we expect to be the outcome of the operation. We define the gate fidelity through the trace distance between the matrices as

\be
\label{eq6}
F=1-\frac{1}{2}\text{Tr}\sqrt{\left(\chi _{id}-\tilde {\chi }\right)^+\left(\chi_{id}-\tilde {\chi }\right)}.
\ee

\noindent The gate error is expressed as  $1-F$.

\subsection{Single-qubit gates} 

\begin{figure}[!t]
 \center
\includegraphics[width=\columnwidth]{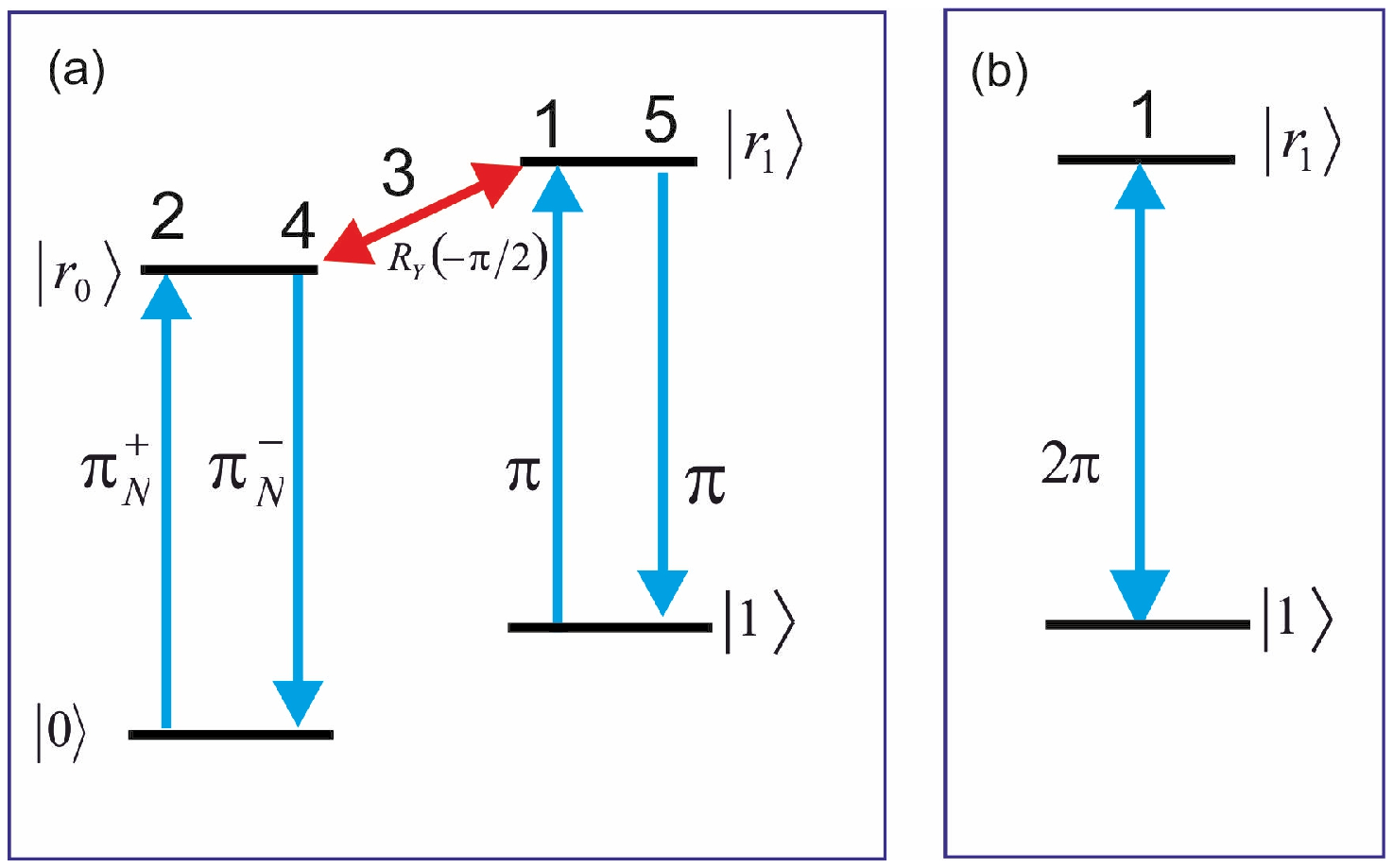}
\vspace{-.5cm}
\caption{
\label{Hadamard}(Color online).
(a) Scheme of the Hadamard gate with mesoscopic atomic ensemble.  Pulses 1-5 act between the qubit states $\ket{0}$, $\ket{1}$ and the Rydberg states $r_0$ and $r_1$. Pulses 2 and 4 are two-photon STIRAP sequences with opposite signs of the detuning from the intermediate state. Pulses~1 and 5 are coherent single-atom $\pi$ pulses. Pulse~3 is a microwave or Raman transition between Rydberg  $r_0$ and $r_1$ with the area $\pi/2$ and phase $\pi/2$. Only one Rydberg excitation in the ensemble is allowed due to Rydberg blockade.  (b) Scheme of the NOT-Z gate with a mesoscopic atomic ensemble. Pulse 1 is a coherent $2\pi$ pulse between the qubit state $\ket{1}$ and Rydberg state $r_1$.
}
\end{figure}

We have numerically studied the fidelity of the quantum-state preparation and single-qubit gates NOT-X, NOT-Y, NOT-Z and Hadamard gate with mesoscopic atomic ensembles. The schemes of NOT-Z gate and Hadamard gate are shown in figure~\ref{Hadamard}. An Hadamard gate is a single-qubit  $R_Y\left(-\pi /2\right)$ rotation combined with a NOT-Z gate, which is produced by two  $\pi $ pulses~1 and 5, acting as a $2\pi $ pulse.

Initially, we have prepared the ensemble into the basis states  $\rho _H=\left(\begin{matrix}1&0\\0&0\end{matrix}\right)$;  $\rho _V=\left(\begin{matrix}0&0\\0&1\end{matrix}\right)$;  $\rho _D=\frac 1 2\left(\begin{matrix}1&1\\1&1\end{matrix}\right)$ and  $\rho _R=\frac 1 2\left(\begin{matrix}1&-i\\i&1\end{matrix}\right)$ by applying single-qubit rotations of the initial ground state of the ensemble. Then we have simulated the single-qubit gates and the X and Y rotations required for quantum state tomography. The probabilities $P_0$ to find the ensemble in the ground state and $P_1$ to find a single atom in the ensemble in the state  $|1\rangle $ have been calculated. The  $\tilde {\chi}$-matrix is reconstructed using a maximum likelihood approximation, and the gate fidelity was finally calculated using equation~(\ref{eq6}). 

\begin{figure}[!t]
 \center
\includegraphics[width=\columnwidth]{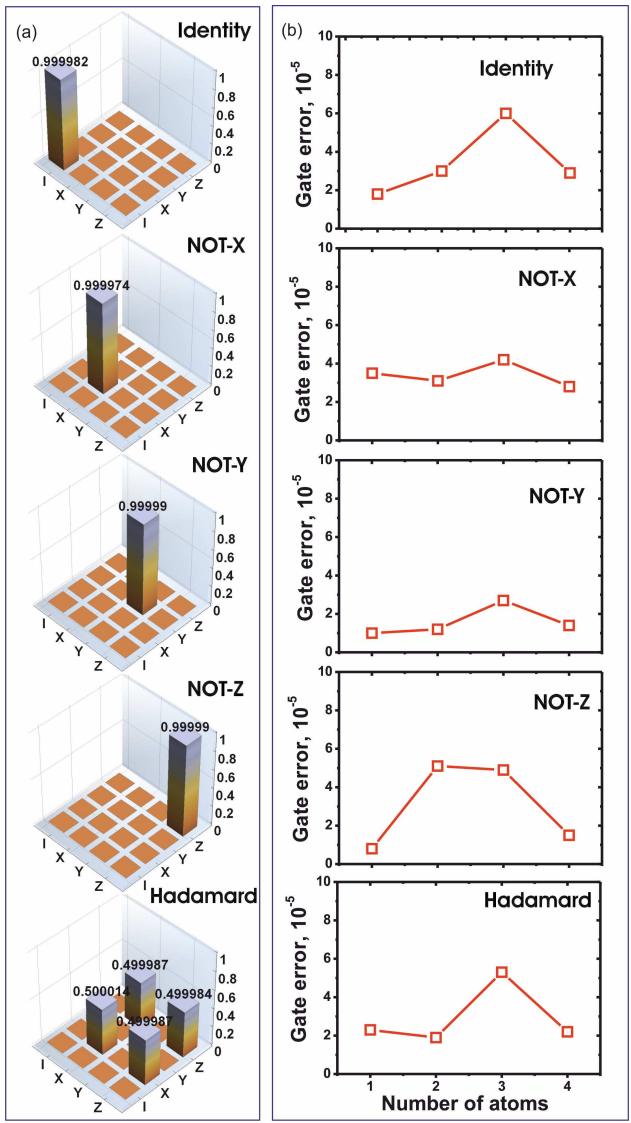}
\vspace{-.5cm}
\caption{
\label{SingleTomo}(Color online).
(a) Numerically simulated  $\tilde{\chi}$-matrices for single-qubit gates (Identity, NOT-X, NOT-Y, NOT-Z and Hadamard) with \textit{N}=4 atoms in the ensemble; (b) Dependences of numerically calculated gate errors of single-qubit gates on the number of atoms in the ensemble.
}
\end{figure}

The reconstructed  $\tilde {\chi}$-matrices for initial state preparation, NOT-X, NOT-Y, NOT-Z and Hadamard gates are presented in left panel of figure~\ref{SingleTomo} for an atomic ensemble with \textit{N}=4 atoms. The gate errors of the single-qubit gates, calculated for atomic ensembles with \textit{N}=1-4 atoms, are shown in the right panel of figure~\ref{SingleTomo}. Regardless of the number of atoms in the ensemble, the gate errors below 10\textsuperscript{-4} have been revealed from the simulations. The small variations of the gate fidelity with the number of atoms, which can be seen in figure~\ref{SingleTomo}(b), are not of significant importance for us. 

In realistic experimental conditions the fidelity of the gates can be deteriorated by a number of undesirable effects, which include finite lifetimes of Rydberg and intermediate excited states, finite Rydberg interaction strength which can lead to a blockade breakdown and destruction of the coherence, fluctuations of laser frequency and intensity. The influence of these effects has been discussed in our previous papers~\cite{Beterov2013,Beterov2014, Beterov2014a}. We expect that the obtained values of the single-qubit fidelity are close to the upper limit which requires precise control of the experimental conditions. We also believe that these simulations confirm the validity of the schemes of quantum logic gates based on adiabatic passage and Rydberg blockade.

\subsection{Two-qubit gate}

A two-qubit process tomography is required for complete reconstruction of two-qubit operations. It includes quantum state tomography of 16 bipartite states of two-qubit systems which is extremely time-consuming for mesoscopic ensembles of multilevel atoms. The fidelity of a CNOT gate can be estimated by measurement of the fidelity of the Bell states, which are created by Hadamard gate applied to a control qubit, and a subsequent CNOT applied to a pair of qubits. 

The Bell states of a bipartite quantum system are defined as following:

\bea
\label{eq7}
\Phi^+&=&\frac 1{\sqrt 2}\left(|00\rangle +|11\rangle \right) \nonumber\\ 
\Phi ^-&=&\frac 1{\sqrt 2}\left(|00\rangle -|11\rangle \right) \nonumber\\
\Psi ^+&=&\frac 1{\sqrt 2}\left(|01\rangle +|10\rangle \right) \nonumber\\
\Psi ^-&=&\frac 1{\sqrt 2}\left(|01\rangle -|10\rangle \right).
\eea

\noindent We have simulated generation of the Bell states as following sequence:

\begin{enumerate}
\item Preparation of two ensemble qubits into the states  $|\bar 0\bar 0\rangle $,  $|\bar 0\bar 1\rangle $,  $|\bar 1\bar 0\rangle $ and  $|\bar 1\bar 1\rangle $ by  $R_y\left(\pi /2\right)$ rotations of the control and target qubits. Both qubits are initially in the state  $|\bar 0\bar 0\rangle $.
\item Single-qubit Hadamard gate with a control ensemble qubit.
\item CNOT-type gate as shown in figure~\ref{Rotation}(b)
\item Quantum state tomography of the final state of two-qubit system.
\end{enumerate}

\begin{figure}[!t]
 \center
\includegraphics[width=\columnwidth]{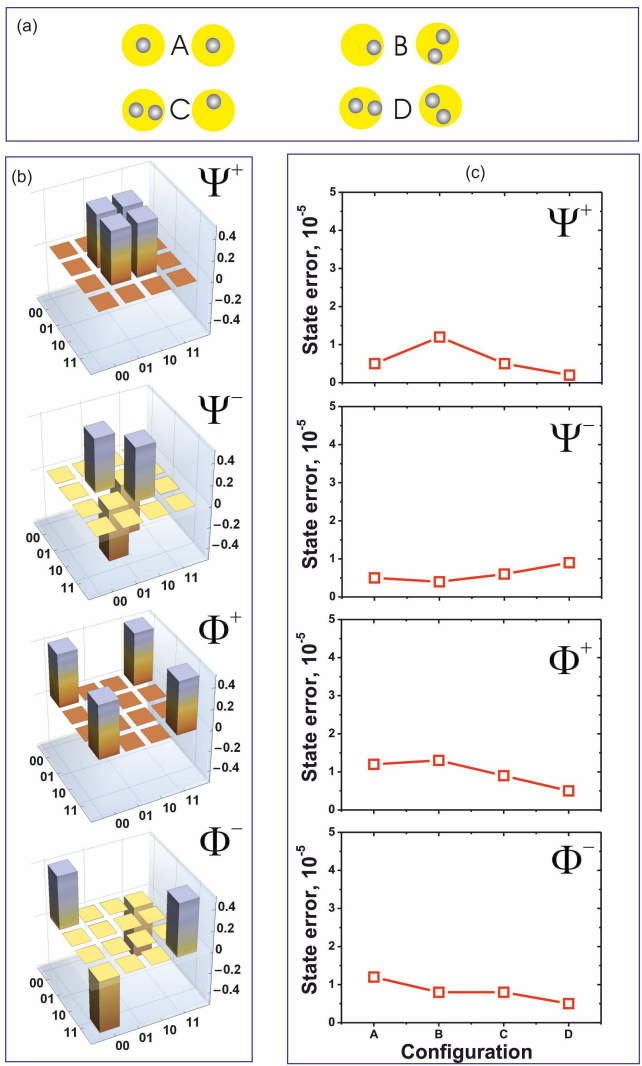}
\vspace{-.5cm}
\caption{
\label{Bell}(Color online).
(a) Scheme of configurations of two interacting mesoscopic ensembles used in the simulation; (b) The reconstructed density matrices of the Bell states; (c) The calculated errors of Bell states for different configurations of two interacting atomic ensembles.
}
\end{figure}

The density matrices of the generated Bell states after using maximum-likelihood reconstruction are shown in figure~\ref{Bell}.  The fidelity of the Bell states has been calculated for \textit{N}=2-4 interacting atoms in the following spatial configurations, shown in figure~\ref{Bell}(a): (A) both control and target ensemble contain a single atom; (B) control ensemble contains one atom and target ensemble contains two atoms; (C) control ensemble contains two atoms and target ensemble contains one atom; (D) both control and target ensemble contain two atoms. High fidelity of the state preparation has been revealed for all Bell states regardless of the configuration of interacting ensembles. This ensures that the infidelity of the CNOT-like gate is kept below 10\textsuperscript{-4}, as it is required for quantum computing. 

\begin{figure}[!t]
 \center
\includegraphics[width=\columnwidth]{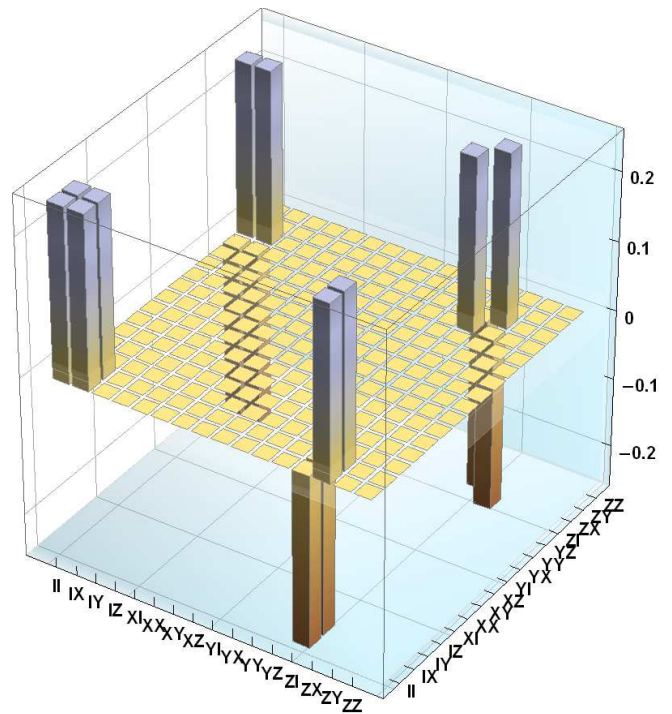}
\vspace{-.5cm}
\caption{
\label{CNOT}(Color online).
The numerically simulated  $\tilde{\chi}$-matrix for a CNOT-type gate for two interacting atoms. The scheme of the gate is shown in figure~\ref{Rotation}(b).
}
\end{figure}

For the sake of completeness, we have simulated a two-qubit quantum process tomography of the CNOT-type gate for the simplest case of two interacting atoms [case (A) in figure~\ref{Bell}(a)]. A 7-pulse sequence for a CNOT-type gate, shown in figure~\ref{Rotation}(b), was used in the simulation. The reconstructed  $\tilde {\chi }$-matrix is shown in figure~\ref{CNOT}. The calculated gate error is below  $4\times 10^{-5}$.

\section{Error sources}
Our simulations have demonstrated high fidelity of the gates, below 10\textsuperscript{-4} for single-qubit gates and for generation of two-qubit Bell states. However, in real experiments a number of error sources may increase the gate errors. The most important limiting factors for quantum computing with Rydberg atoms are the following:
\begin{enumerate}
\item \textit{Rydberg blockade breakdown}. 
The atoms must be prepared in tightly focused optical dipole traps to ensure the regime of perfect Rydberg blockade~\cite{Beterov2014a}. Recent experiments~\cite{Ebert2015} have demonstrated the coherence of ensemble qubit states and a strong Rydberg blockade between spatially separated ensembles. Quantum gates and entanglement of two ensemble qubits have not yet been demonstrated. One issue is that the atomic interactions in the ensemble of multilevel atoms with variable spacings and interactions strengths can lead to dephasing and blockade breakdown~\cite{Derevianko2015}, this problem requires further investigation.
\item \textit{Finite lifetime of the Rydberg state}. Rydberg atoms with $n\sim100$ have long room-temperature lifetimes of around 200 microseconds~\cite{Beterov2009}. However, the decay of the Rydberg state during temporary Rydberg excitation substantially reduces the gate fidelity. This effect can be suppressed by reducing the interval when the atom is kept in the Rydberg state, but at a price of higher Rabi frequencies which requires higher laser powers. 
\item \textit {Finite lifetime of the intermediate excited state}. The first excited alkali-metal states typically have short lifetimes of tens of nanoseconds. Spontaneous decay of these states destroys the coherence of multi-photon excitation. This problem can be partly avoided by an increase of the detuning from the intermediate excited state for two-photon excitation. 
\item \textit{Laser intensity fluctuations}. The scheme which we propose is sensitive to asymmetry of the pulses in double STIRAP sequence~\cite{Beterov2013, Beterov2014}, but is much less sensitive to small variations of the Rabi frequency for different double STIRAP sequences which can be caused by slow changes of the laser intensity between subsequent gates.  
\item \textit{Finite temperature of the atoms in the trap}. In our simulations we assumed zero temperature of the trapped atoms (frozen Rydberg gas). The Doppler shift due to the finite temperature of the atoms may result in slightly detuned Rydberg excitation leading to the errors below $10^{-5}$ at temperatures $100\,\mu$K~\cite{Saffman2011}. Another problem is dephasing of the Rydberg state relative to the ground state during gate operation. A detailed analysis of this effect is given in Ref.~\cite{Saffman2011}.
\item \textit{Dephasing of the collective state of the superatom}. The fluctuations of the phases of the lasers and the spatial variation of the laser intensity may lead both to homogeneous and inhomogeneous dephasing of the collective states of the atomic ensembles containing randomly distributed atoms~\cite{Honer2011}. 
\end{enumerate}
Strong Rydberg-Rydberg interaction is required to achieve the regime of perfect Rydberg blockade both within the atomic ensemble and for two interacting ensembles. Simultaneous excitation of two atoms into the states $\ket{r_0r_0}$, $\ket{r_1r_1}$ or $\ket{r_0r_1}$ must be completely suppressed. 
The long range interaction strength can be parameterized with a $C_6$ coefficient as $V(n,n')=C_6^{(n,n')}/R^6$ with $R$ the atomic separation.  For Cs $nS$ states the optimum gate fidelity is obtained for $80S$~\cite{HZhang2012}, and the interaction strengths for $\ket{r_0}=\ket{80S_{1/2},m=1/2}, \ket{r_1}=\ket{81S_{1/2},m=1/2}$ are 
$C_6^{(80,80)}=3.2$, $C_6^{(80,81)}=5.1$, $C_6^{(81,81)}=3.7$, in units of $10^6\, {\rm  MHz}\, \mu {\rm m}^6$.  Rydberg $nS$ states can be accessed starting from a ground $S$ state using two-photon STIRAP pulses. Although the interaction of $nP$ states is not isotropic it can be made isotropic in 1- or 2-dimensional lattices by orienting the quantization axis perpendicular to the lattice symmetry plane. For Cs atoms the optimal state 
is  $112P_{3/2}$~\cite{HZhang2012}, and the interaction strengths for $\ket{r_0}=\ket{112P_{3/2},m=3/2}, \ket{r_1}=\ket{113p_{3/2},m=3/2}$ at 90 deg. to the quantization axis are 
$C_6^{(112,112)}=250$, $C_6^{(112,113)}=820$, $C_6^{(113,113)}=270$, in units of $10^6\, {\rm  MHz}\, \mu {\rm m}^6$. For both $nS$ and $nP$ states a strong interaction is obtained for all involved  Rydberg states as desired.
The control over the interaction strength using rf-assisted F\"orster resonances~\cite{Tretyakov2014} can be also of interest.

The pulse connecting $\ket{r_0}, \ket{r_1}$ can be implemented as a 2-photon electric dipole transition at microwave frequencies via a neighboring opposite parity state or as a two-photon laser Raman transition. The large transition dipole moments of Rydberg states scaling as $n^2$ render fast microwave pulses straightforward to implement. At $n=80$, a detuning of 1~GHz from the intermediate state, and a small $1~\mu\rm W/cm^2$ microwave power level, gives $\sim 25~\rm MHz$ two-photon Rabi frequency. 

\begin{figure}[!t]
\center
\includegraphics[width=0.8\columnwidth]{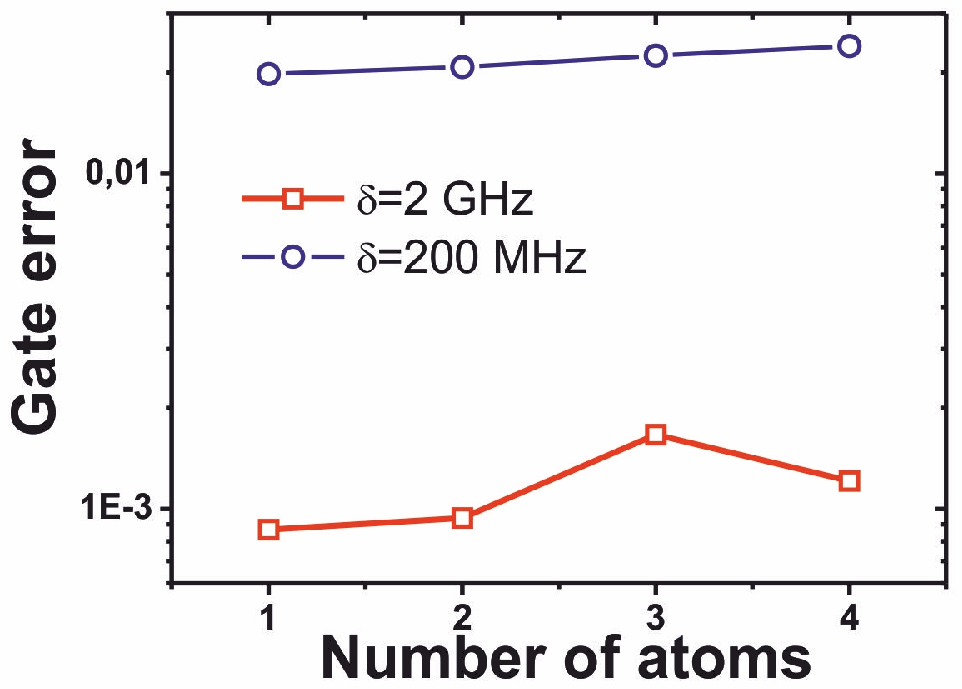}
\vspace{-.5cm}
\caption{
\label{Err}
The calculated dependence of the error of population transfer on the number of atoms taking into account finite linewidths of the intermediate excited state $\gamma / (2\pi)=5$~MHz and Rydberg state $\gamma_R / (2\pi)=0.8$~kHz. Circles: $T_0=2$~$\mu$s, $\delta / (2\pi)=$200~MHz, and $\Omega / (2\pi)=$50~MHz. Squares: $T_0=100$~ns, $\delta / (2\pi)=$2~GHz, and $\Omega /(2\pi)=$500~MHz.
}
\end{figure}

To estimate the effect of the finite lifetimes of the intermediate excited and Rydberg states on the gate fidelities we have simulated a STIRAP in an atomic ensemble using the master equation~\cite{Petrosyan2013,Beterov2013}. The calculated population error after the first STIRAP sequence with the parameters from figure~\ref{Scheme} is substantially increased if linewidth of the intermediate state $\gamma / (2\pi)=5$~MHz and decay of the Rydberg state with $\gamma_R / (2\pi)=0.8$~kHz  are taken into account, as shown in figure~\ref{Err} (circles).

To reduce this effect, we considered short pulses with large Rabi frequencies and detunings from the intermediate state. 
We have taken $T_0=100$~ns, $\delta / (2\pi)=$2~GHz, and $\Omega / (2\pi)=$500~MHz. The calculated population errors are shown in figure~\ref{Err} for \textit{N}=1-4 atoms. Although the error exceeds $10^{-4}$, which is required for quantum error correction, it is still smaller than $2\times 10^{-3}$ regardless of the number of atoms.

We have simulated the quantum process tomography of the Hadamard gate taking into account finite lifetimes of the intermediate and Rydberg states using a master equation for the density matrix in the conditions of figure~\ref{Err}. The atomic ensembles with a small number of atoms \textit{N}=1 and \textit{N}=2 were considered. For the small detuning from the intermediate state $\delta / (2\pi)=$200~MHz the calculated error was higher than 10\%. For the increased detuning $\delta / (2\pi)=$2000~MHz  and reduced time interval between the laser pulses ($t_1=-170$~ns and $t_2=170$~ns) the calculated error is 0.4\% for \textit{N}=1 and 2.1\% for \textit{N}=2. The time interval between the pulses 1 and 5 from figure~\ref{Rotation} was reduced to 600~ns. This error includes finite accuracy of the state preparation and measurement. Quantum gate error can be further reduced by increasing the laser intensities and detuning from the intermediate state along with excitation of Rydberg states with larger lifetimes and shortening  the time intervals between the laser pulses~\cite{Saffman2010}.

\section{Summary}

The simulated single-qubit and two-qubit quantum process tomography confirms usability of the quantum gates based on adiabatic passage and Rydberg blockade with mesoscopic atomic ensembles. High fidelity of the gates required for quantum computing can be achieved by use of optimized shapes of the STIRAP pulses. The gate error has been found to be below 10\textsuperscript{-4 }for single-qubit gates and for generation of two-qubit Bell states. For experimental implementation, as it has been shown in our previous works~\cite{Beterov2013, Beterov2014, Beterov2014a} it would be necessary to increase the detuning from the intermediate excited state up to 2 GHz~\cite{Beterov2013} to reduce the effect of its short lifetime. The proposed scheme of the quantum gates is insensitive to the exact value of Rabi frequency of STIRAP pulses, provided the adiabaticity condition is fulfilled, but is sensitive to asymmetry of the pulses in the STIRAP sequence~\cite{Beterov2013}. Atoms must be prepared in tightly focused optical dipole traps to ensure the regime of perfect Rydberg blockade~\cite{Beterov2014a}.

\section*{Acknowledgements}
This work was supported by Russian Science Foundation Grant No. 16-12-00028 in part of simulation of Bell states and RFBR Grant No. 14-02-00680, Novosibirsk State University and Russian Academy of Sciences.  MS was supported by NSF award 1521374, the AFOSR MURI on Quantum Memories and Light-Matter Interfaces, and the ARL-CDQI through cooperative agreement W911NF-15-2- 0061.

\appendix

\section*{Appendix: quantum tomography of single-qubit and two-qubit gates in a two-level system using Rabi rotations}

\subsection{Interaction of a two-level qubit with resonant laser radiation and rotations on a Bloch sphere}

The quantum state of a two-level qubit can be written as

\be
\label{eqA1}
c_0|0\rangle +c_1|1\rangle =\left[\text{cos}\left(\frac{\theta _0} 2\right)|0\rangle +e^{\mathit{i\varphi }_0}\text{sin}\left(\frac{\theta _0} 2\right)|1\rangle \right]e^{\mathit{i\gamma}}.
\ee

\noindent Here the angles  $\theta _0$ and  $\varphi _0$ define the position of the qubit on a Bloch sphere and  $\gamma $ is an unimportant phase factor which can be omitted. The interaction with resonant laser radiation is described by the system of two differential equations for the probability amplitudes: 

\be
\label{eqA2}
i\left(\begin{matrix}\dot c_0\\\dot c_1\end{matrix}\right)=\frac 1 2\left(\begin{matrix}0&\Omega ^{\ast }\\
\Omega &0\end{matrix}\right)\left(\begin{matrix}c_0\\c_1\end{matrix}\right).
\ee

\noindent Here  $\Omega =\Omega _0e^{\mathit{i\varphi }}$ is a complex Rabi frequency which takes into account the phase of the laser field. \ The solution of the system~(\ref{eqA2}) is expressed as a Rabi rotation of the initial vector state: 

\be
\label{eqA3}
\begin{pmatrix}c'_0\\c'_1\end{pmatrix}=
\begin{pmatrix}\text{cos}\frac{\theta}{2}& i\text{e}^{-\mathit{i\varphi }}\text{sin}\frac{\theta }{2}\\
i\text{e}^{\mathit{i\varphi}}\text{sin}\frac{\theta}{2}&\text{cos}\frac{\theta}{2}\end{pmatrix}\begin{pmatrix}c_0\\c_1\end{pmatrix}=R\left(\theta ,\varphi \right)\begin{pmatrix}c_0\\c_1\end{pmatrix}.
\ee

\noindent Here  $\theta =-\Omega _0T$, where \textit{T} is the time duration of interaction of the qubit with laser radiation.

The qubit rotations are described by the rotation matrices:

\bea
\label{eqA4}
R_X\left(\theta \right)=\text{exp}\left(-i\frac{\theta } 2\sigma_x\right)=\begin{pmatrix}\text{cos}\frac{\theta}{2} &-i\text{sin}\frac{\theta}{2} \\-i\text{sin}\frac{\theta}{2} &\text{cos}\frac{\theta}{2} \end{pmatrix} \nonumber \\
R_Y\left(\theta \right)=\text{exp}\left(-i\frac{\theta}{2}\sigma_y\right)=\begin{pmatrix}\text{cos}\frac{\theta}{2} &-\text{sin}\frac{\theta}{2} \\\text{sin}\frac{\theta}{2} &\text{cos}\frac{\theta }{2}\end{pmatrix}\nonumber \\
R_Z\left(\theta \right)=\text{exp}\left(-i\frac{\theta}{2}\sigma_z\right)=\begin{pmatrix}e^{-\mathit{i\theta}/2} &0\\0 &e^{\mathit{i\theta}/2}
\end{pmatrix}.
 \eea

\noindent For X and Y rotations from equation~(\ref{eqA4}) we find  $R_x\left(\theta \right)=R\left(-\theta ,0\right)$ and  $R_y\left(\theta \right)=R\left(-\theta ,\pi /2\right)$.

We can prepare the initial single-qubit states  $\rho _H=\left(\begin{matrix}1&0\\0&0\end{matrix}\right)$;  $\rho _V=\left(\begin{matrix}0&0\\0&1\end{matrix}\right)$;  $\rho _D=\frac 1 2\left(\begin{matrix}1&1\\1&1\end{matrix}\right)$ and  $\rho _R=\frac 1 2\left(\begin{matrix}1&-i\\i&1\end{matrix}\right)$ starting from  $\rho _H$ by single-qubit rotations:

\bea
\label{eqA5}
\rho_V&=&R_Y\left(\pi \right)\rho_HR_Y^{\dag}\left(\pi \right) \nonumber \\
\rho_D&=&R_Y\left(\pi /2\right)\rho _HR_Y^{\dag}\left(\pi /2\right) \nonumber \\
\rho_R&=&R_X\left(-\pi /2\right)\rho _HR_X^{\dag}\left(-\pi /2\right).
\eea

\subsection{Single-qubit state tomography} 
A single-qubit density matrix can be written as  $\rho _{\left(1\right)}=\frac{1}{2}\sum\limits_{i=1}^{4}\text{Tr}\left(\sigma _i\rho _{\left(1\right)}\right)\sigma _i$ where  $\sigma _i$ are four Pauli matrices \textit{I},  $\sigma _x$, $\sigma _y$ and  $\sigma _z$. That means that we can express the quantum state through four quantum mechanical observables. Two of them can be obtained by a measurement of the probabilities $P_0$ and $P_1$ to find a qubit in the state  $\ket{0}$ or  $\ket{1}$:

\bea
\label{eqA6}
\lambda _1&=&\text{Tr}\left(\sigma _0\rho_{\left(1\right)}\right)=\rho _{00}+\rho _{11}=P_0+P_1 \nonumber  \\
\lambda _4&=&\text{Tr}\left(\sigma _z\rho _{\left(1\right)}\right)=\rho _{00}-\rho _{11}=P_0-P_1.
\eea

\noindent The other observables can be expressed through the probabilities $P_0$ and $P_1$ to find a qubit in the state  $\ket{0} $ or  $\ket{1} $ after single-qubit rotations around X and Y axes. From the expressions 

\bea
\label{eqA7}
\lambda _2&=&\text{Tr}\left(\sigma _x\rho _{\left(1\right)}\right)=\text{Tr}\left(\sigma _zR_y\left(-\pi /2\right)\rho _{\left(1\right)}R_y^{\dag}\left(-\pi /2\right)\right)\nonumber \\
\lambda_3&=&\text{Tr}\left(\sigma _y\rho _{\left(1\right)}\right)=\text{Tr}\left(\sigma _zR_x\left(\pi /2\right)\rho _{\left(1\right)}R_x^{\dag}\left(\pi /2\right)\right).
\eea

\noindent we find that after  $R_y\left(-\pi /2\right)$ and  $R_x\left(\pi /2\right)$ \ \ rotations we should measure the values  $\lambda _{2,3}=P_0-P_1$ and reconstruct the density matrix as \  $\rho _{\left(1\right)}=\frac 1 2\overset 4{\underset{i=1}{\sum }}\lambda _i\sigma _i$.

The equations~(\ref{eqA6}) and (\ref{eqA7}) can be presented in a table form:

\begin{table*}
\caption{Single-qubit tomography}
\begin{tabular*}{\textwidth}{@{\extracolsep{\fill}}|c|c|c|}  \hline
coefficient & Action on qubit & Measured value\\\hline
$\lambda_1$ &{\itshape I} & $P_0+P_1$\\\hline
$\lambda _2$ & $R_y\left(-\pi /2\right)$ & $P_0-P_1$\\\hline
$\lambda _3$ & $R_x\left(\pi /2\right)$ & $P_0-P_1$\\\hline
$\lambda _4$ &{\itshape I} & $P_0-P_1$\\\hline
\end{tabular*}
\end{table*}
\subsection{Two-qubit state tomography} 

A two-qubit density matrix is written as:

\be
\label{eqA8}
\rho _{\left(2\right)}=\frac {1}{4}\sum\limits_{i,j=1}^{4}\text{Tr}\left[\left(\sigma_i{\otimes}\sigma_j\right)\rho_{\left(2\right)}\right]\left(\sigma _i{\otimes}\sigma_j\right).
\ee

\noindent  We can reconstruct it as 
\be
\label{eqA9}
\rho _{\left(1\right)}=\frac{1}{4}\sum\limits_{i,j=1}^{4}\lambda_{ij}\left(\sigma _i{\otimes}\sigma_j\right).
\ee

\noindent The coefficients  $\lambda_{ij}$ are expressed through the measured probabilities  $P_{00}$,  $P_{01}$,  $P_{10}$,  $P_{11}$ to find the bipartite system in states $\ket{00} $,  $\ket{01}$,  $\ket{10}$ and  $\ket{11}$, respectively. The sequence of measurements required to find $\lambda_{ij}$ is presented in table~2.
\begin{table*}
\caption{Two-qubit state tomography}
\begin{tabular*}{\textwidth}{@{\extracolsep{\fill}}|c|c|c|c|}  \hline
coefficient & Action on control qubit & Action on target qubit & Measured value\\\hline
 $\lambda _{11}$ &{\itshape I} &{\itshape I} & $P_{00}+P_{01}+P_{10}+P_{11}$\\\hline
 $\lambda _{12}$ &{\itshape I} & $R_y\left(-\pi /2\right)$ & $P_{00}-P_{01}+P_{10}-P_{11}$\\\hline
 $\lambda _{13}$ &{\itshape I} & $R_x\left(\pi /2\right)$ & $P_{00}-P_{01}+P_{10}-P_{11}$\\\hline
 $\lambda _{14}$ &{\itshape I} &{\itshape I} & $P_{00}-P_{01}+P_{10}-P_{11}$\\\hline
 $\lambda _{21}$ & $R_y\left(-\pi /2\right)$ &{\itshape I} & $P_{00}+P_{01}-P_{10}-P_{11}$\\\hline
 $\lambda _{22}$ & $R_y\left(-\pi /2\right)$ & $R_y\left(-\pi /2\right)$ & $P_{00}-P_{01}-P_{10}+P_{11}$\\\hline
 $\lambda _{23}$ & $R_y\left(-\pi /2\right)$ & $R_x\left(\pi /2\right)$ & $P_{00}-P_{01}-P_{10}+P_{11}$\\\hline
 $\lambda _{24}$ & $R_y\left(-\pi /2\right)$ &{\itshape I} & $P_{00}-P_{01}-P_{10}+P_{11}$\\\hline
 $\lambda _{31}$ & $R_x\left(\pi /2\right)$ &{\itshape I} & $P_{00}+P_{01}-P_{10}-P_{11}$\\\hline
 $\lambda _{32}$ & $R_x\left(\pi /2\right)$ & $R_y\left(-\pi /2\right)$ & $P_{00}-P_{01}-P_{10}+P_{11}$\\\hline
 $\lambda _{33}$ & $R_x\left(\pi /2\right)$ & $R_x\left(\pi /2\right)$ & $P_{00}-P_{01}-P_{10}+P_{11}$\\\hline
 $\lambda _{34}$ & $R_x\left(\pi /2\right)$ &{\itshape I} & $P_{00}-P_{01}-P_{10}+P_{11}$\\\hline
 $\lambda _{41}$ &{\itshape I} &{\itshape I} & $P_{00}+P_{01}-P_{10}-P_{11}$\\\hline
 $\lambda _{42}$ &{\itshape I} & $R_y\left(-\pi /2\right)$ & $P_{00}-P_{01}-P_{10}+P_{11}$\\\hline
 $\lambda _{43}$ &{\itshape I} & $R_x\left(\pi /2\right)$ & $P_{00}-P_{01}-P_{10}+P_{11}$\\\hline
 $\lambda _{44}$ &{\itshape I} &{\itshape I} & $P_{00}-P_{01}-P_{10}+P_{11}$\\\hline
\end{tabular*}
\end{table*}

\subsection{Single-qubit process tomography}
To perform a single-qubit process tomography, we select an operator basis  $\tilde  E_i=\sigma _i$ with four Pauli matrices \textit{I},  $\sigma _x$, $\sigma _y$ and  $\sigma _z$. The action of the unitary quantum gate \textit{U} on the density matrix of the initial state is expressed as 

\be
\label{eqA10}
 \rho'=\varepsilon \left(\rho \right)=U\mathit{\rho U^{\dag}}.
\ee

\noindent For the basis states  $\rho _1=\left(\begin{matrix}1&0\\0&0\end{matrix}\right)$,  $\rho _2=\left(\begin{matrix}0&1\\0&0\end{matrix}\right)$,  $\rho _3=\left(\begin{matrix}0&0\\1&0\end{matrix}\right)$, and  $\rho _4=\left(\begin{matrix}0&0\\0&1\end{matrix}\right)$ it has been shown that the  $\chi$-matrix can be reconstructed through the block matrix built of the density matrices of the quantum states, measured after the performed quantum gate~\cite{Nielsen2011}: 

\be
\label{eqA11}
\chi =\Lambda \left(\begin{matrix}\rho'_1&\rho'_2\\\rho'_3&\rho'_4\end{matrix}\right)\Lambda.
\ee

\noindent Here the block matrix  $\Lambda =\frac 1 2\left(\begin{matrix}I&\sigma _x\\\sigma _x&-I\end{matrix}\right)$ .

In the experiment we prepare the qubit into the basis states  $\rho _H$, $\rho _V$, $\rho _D$, $\rho _R$ and finally get the states  $\rho'_H$, $\rho'_V$, $\rho'_D$, $\rho'_R$ after the gate operation. To use equation~(\ref{eqA11}), we need to find the matrices $\rho'_1$, $\rho'_2$, $\rho'_3$, $\rho'_4$ through the following transformation: 

\be
\label{eqA12}
\left(\begin{matrix}\rho'_1\\\rho_2'\\\rho'_3\\\rho'_4\end{matrix}\right)=\left(\begin{matrix}1&0&0&0\\-a&-a&1&i\\-a^{\ast }&-a^{\ast }&1&-i\\0&1&0&0\end{matrix}\right)\left(\begin{matrix}\rho'_H\\\rho'_V\\\rho' _D\\\rho'_R\end{matrix}\right).
\ee

\noindent Here  $a=\frac 1 2\left(1+i\right)$.
\subsection{Two-qubit process tomography} 

For two-qubit process tomography, we select the operator basis  $\tilde  E_{4(i-1)+j}=\sigma _i{\otimes}\sigma _j$ with $i,j=1-4$. The basis states  $\rho _{ij}$ are matrices with~1 at $\mathrm{i}_{th}$ row and $\mathrm{j}_{th}$column. The  $\chi$-matrix is reconstructed using a block matrix of the measured density matrices~\cite{Nielsen2011, White2007}:

\be
\label{eqA13}
\chi =K^T\left(\begin{matrix}
\rho'_{11}&\rho'_{12}&\rho'_{13}&\rho'_{14}\\
\rho'_{21}&\rho'_{22}&\rho'_{23}&\rho'_{24}\\
\rho'_{31}&\rho'_{32}&\rho'_{33}&\rho'_{34}\\
\rho'_{41}&\rho'_{42}&\rho'_{43}&\rho'_{44}
\end{matrix}\right)K.
\ee

\noindent Here  $K=\mathit{P\Lambda }$,  $P=I{\otimes}\left[M{\otimes}I\right]$,  $\Lambda =\frac 1 4\left(\sigma _Z{\otimes}I+\sigma _X{\otimes}\sigma _X\right){\otimes}\left(\sigma _Z{\otimes}I+\sigma _X{\otimes}\sigma _X\right)$ and

\be
\label{eqA14}
M=\left(\begin{matrix}1&0&0&0\\0&0&1&0\\0&1&0&0\\0&0&0&1\end{matrix}\right).
\ee

\noindent Similarly to a single-qubit tomography, in the experiment we prepare two qubits in the bipartite physical basis  $\rho_{AB}$ with  $A,B=\left\{H,V,D,R\right\}$. To use equation~(\ref{eqA13}),  it is necessary to find the density matrices  $\rho'_{ij}$ after measurement of the final states  $\rho'_{AB}$ of two qubits by the following transformation~\cite{White2007}:
\begin{widetext}
\begin{equation}
\label{eqA15}
\left(\begin{smallmatrix}\rho'_{11}\\\rho'_{12}\\\rho'_{13}\\\rho'_{14}\\\rho'_{21}\\\rho'_{22}\\\rho'_{23}\\\rho'_{24}\\\rho'_{31}\\\rho'_{32}
\\\rho'_{33}\\\rho'_{34}\\\rho'_{41}\\\rho'_{42}\\\rho'_{43}\\\rho'_{44}\end{smallmatrix}\right)=
\left(\begin{smallmatrix}
1&0&0&0&0&0&0&0&0&0&0&0&0&0&0&0\\
-a&-a&1&i&0&0&0&0&0&0&0&0&0&0&0&0\\
-a&0&0&0&-a&0&0&0&1&0&0&0&i&0&0&0\\
i/2&i/2&-a&a^{\ast}&i/2&i/2&-a&a^{\ast}&-a&-a&1&i&a^{\ast }&a^{\ast }&i&-1\\
-a^{\ast }&-a^{\ast}&1&-i&0&0&0&0&0&0&0&0&0&0&0&0\\
0&1&0&0&0&0&0&0&0&0&0&0&0&0&0&0\\
1/2&1/2&-a&-a^{\ast}&1/2&1/2&-a&-a^{\ast}&-a^{\ast }&-a^{\ast}&1&-i&-a&-a&i&1
\\0&-a&0&0&0&-a&0&0&0&1&0&0&0&i&0&0\\-a^{\ast }&0&0&0&-a^{\ast}&0&0&0&1&0&0&0&-i&0&0&0\\
1/2&1/2&-a^{\ast }&-a&1/2&1/2&-a^{\ast }&-a&-a&-a&1&i&-a^{\ast}&-a^{\ast}&-i&1\\
0&0&0&1&0&0&0&0&0&0&0&0&0&0&0&0\\
0&0&0&-a&-a&1&i&0&0&0&0&0&0&0&0&0\\
-i/2&-i/2&-a^{\ast}&a&-i/2&-i/2&-a^{\ast}&a&-a^{\ast }&-a^{\ast}&1&-i&a&a&-i&-1\\
0&-a^{\ast }&0&0&0&-a^{\ast}&0&0&0&1&0&0&0&-i&0&0\\
0&0&0&0&-a^{\ast }&-a^{\ast}&1&-i&0&0&0&0&0&0&0&0\\
0&0&0&0&0&1&0&0&0&0&0&0&0&0&0&0
\end{smallmatrix}\right)=\left(\begin{smallmatrix}\rho'_{HH}\\\rho'_{HV}\\\rho'_{HD}\\\rho'_{HR}\\\rho'_{VH}\\\rho'_{VV}\\\rho'_{VD}\\\rho'_{VR}\\\rho'_{DH}\\\rho'_{DV}\\\rho'_{DD}
\\\rho'_{DR}\\\rho'_{RH}\\\rho'_{RV}\\\rho'_{RD}\\\rho'_{RR}\end{smallmatrix}\right).
\end{equation}
\end{widetext}

\noindent The quantum process is reconstructed as  $\rho'=\varepsilon \left(\rho \right)=\sum\limits_{i,j}\chi_{ij}\tilde  E_i\rho \tilde  E_j^+$ both for single-qubit and two-qubit tomography.

\subsection{Maximum-likelihood reconstruction} 
The density matrices and the  $\chi$-matrices reconstructed from measurements may be non-physical. The idea of a maximum-likelihood approximation is to find the matrix which is closest to the measured one. Any physical single-qubit density matrix can be written as  $\tilde {\rho }^{\left(1\right)}=T^{\left(1\right)^+}T^{\left(1\right)}$ where 
\be
\label{eqA16}
 T^{\left(1\right)}=\left(\begin{matrix}t_1&0\\t_3+i t_4&t_2\end{matrix}\right).
\ee

\noindent Here  $\vec t=\left\{t_i\right\}$ is a vector of real parameters. To find the density matrix  $\tilde {\rho }^{\left(1\right)}$ which approximates the measured density matrix  $\rho ^{\left(1\right)}$ we find the minimum of the function

\be
\label{eqA17}
 \Delta _{\rho }\left(\vec t\right)=\sum\limits_{m,n=1}^{2}|\tilde {\rho }_{mn}^{\left(1\right)}\left(\vec t\right)-\rho_{mn}^{\left(1\right)}|^2.
\ee

\noindent We keep constraints  $\text{Tr}\left(\tilde {\rho }^{\left(1\right)}\right)=1$ directly in the minimization procedure. Similar approach is used for two-qubit density matrices. 

For reconstruction of the single-qubit  $\chi$-matrix we use a parametrization  $\tilde {\rho }^{\left(1\right)}=T^{\left(2\right)^+}T^{\left(2\right)}$ with

\be
\label{eqA18}
 T^{\left(2\right)}=\left(\begin{matrix}t_1&0&0&0\\t_5+i t_6&t_2&0&0\\t_7+i t_8&t_9+i t_{10}&t_3&0\\t_{11}+i t_{12}&t_{13}+i t_{14}&t_{15}+i t_{16}&t_4\end{matrix}\right).
\ee

\noindent and find minimum of the function

\be
\label{eqA19}
 \Delta _{\chi }\left(\vec t,\lambda \right)=\sum\limits_{m,n=1}^{4}|\tilde {\chi }_{mn}\left(\vec t\right)-\chi_{mn}|^2.
\ee
\noindent We keep 

\be
\label{eqA20}
 \sum\limits_{m,n=1}^{4}\tilde {\chi }_{mn}\left(\vec t\right)\tilde  E_m^+\tilde  E_n=I_4.
\ee

\noindent in the minimization procedure to ensure that the quantum process is trace-preserving. Here $I_4$ is a four-by-four identity matrix.

Equations~(\ref{eqA18}) and (\ref{eqA19}) are easily generalized for two-qubit quantum process tomography, where the vector  $\vec t=\left\{t_i\right\}$ contains 256 components. To reduce computation time, in our simulations we have used constraints~(\ref{eqA20}) only for diagonal elements of the identity matrix.

\subsection{Gate fidelity}

The CNOT-type gate, shown in Fig.~\ref{Rotation}(b), is represented as 

\be
\label{eqA21}
 U_{\text{CNOT-type}}=\begin{pmatrix}0&1&0&0\\1&0&0&0\\0&0&1&0\\0&0&0&1\end{pmatrix}.
\ee

\noindent To estimate the gate fidelity, we first find the ideal process matrix for this CNOT-type gate. The transformation of the density matrices of the basis states is written as  $\rho'_{ij}=U_{\text{CNOT-type}}\rho_{ij}U_{\text{CNOT-type}}^{\dag}$ where  $\rho _{ij}$ are matrices with 1 at $\mathrm{i}_{th}$ row and $\mathrm{j}_{th}$ column. The  $\chi$-matrix is found from equations~(\ref{eqA13}) and (\ref{eqA14}).

\be
\label{eqA22}
\chi_{\text{CNOT-type}}=\frac 1 4 \left(\begin{smallmatrix}1&1&0&0&0&0&0&0&0&0&0&0&-1&1&0&0\\1&1&0&0&0&0&0&0&0&0&0&0&-1&1&0&0\\0&0&0&0&0&0&0&0&0&0&0&0&0&0&0&0\\0&0&0&0&0&0&0&0&0&0&0&0&0&0&0&0\\0&0&0&0&0&0&0&0&0&0&0&0&0&0&0&0\\0&0&0&0&0&0&0&0&0&0&0&0&0&0&0&0\\0&0&0&0&0&0&0&0&0&0&0&0&0&0&0&0\\0&0&0&0&0&0&0&0&0&0&0&0&0&0&0&0\\0&0&0&0&0&0&0&0&0&0&0&0&0&0&0&0\\0&0&0&0&0&0&0&0&0&0&0&0&0&0&0&0\\0&0&0&0&0&0&0&0&0&0&0&0&0&0&0&0\\0&0&0&0&0&0&0&0&0&0&0&0&0&0&0&0\\-1&-1&0&0&0&0&0&0&0&0&0&0&1&-1&0&0\\1&1&0&0&0&0&0&0&0&0&0&0&-1&1&0&0\\0&0&0&0&0&0&0&0&0&0&0&0&0&0&0&0\\0&0&0&0&0&0&0&0&0&0&0&0&0&0&0&0\end{smallmatrix}\right).
\ee

\noindent The fidelity of a quantum gate can be measured by comparison of the reconstructed process matrix with the ideal matrix 

\be
\label{eqA23}
 F=1-\frac 1 2\text{Tr}\sqrt{\left(\chi _{id}-\tilde {\chi }\right)^+\left(\chi _{id}-\tilde {\chi }\right)}.
\ee

\noindent The fidelities of the single-qubit gates have been calculated similarly. To estimate the fidelities of the Bell states, we used a similar expression:

\be
\label{eqA24}
 F=1-\frac 1 2 \text{Tr}\sqrt{\left(\rho_{id}-\tilde {\rho }\right)^+\left(\rho _{id}-\tilde {\rho }\right)}.
\ee

\noindent The gate error is expressed as  $1-F$.


\section*{References}

\begin{thebibliography}{44}%
\makeatletter
\providecommand \@ifxundefined [1]{%
 \@ifx{#1\undefined}
}%
\providecommand \@ifnum [1]{%
 \ifnum #1\expandafter \@firstoftwo
 \else \expandafter \@secondoftwo
 \fi
}%
\providecommand \@ifx [1]{%
 \ifx #1\expandafter \@firstoftwo
 \else \expandafter \@secondoftwo
 \fi
}%
\providecommand \natexlab [1]{#1}%
\providecommand \enquote  [1]{``#1''}%
\providecommand \bibnamefont  [1]{#1}%
\providecommand \bibfnamefont [1]{#1}%
\providecommand \citenamefont [1]{#1}%
\providecommand \href@noop [0]{\@secondoftwo}%
\providecommand \href [0]{\begingroup \@sanitize@url \@href}%
\providecommand \@href[1]{\@@startlink{#1}\@@href}%
\providecommand \@@href[1]{\endgroup#1\@@endlink}%
\providecommand \@sanitize@url [0]{\catcode `\\12\catcode `\$12\catcode
  `\&12\catcode `\#12\catcode `\^12\catcode `\_12\catcode `\%12\relax}%
\providecommand \@@startlink[1]{}%
\providecommand \@@endlink[0]{}%
\providecommand \url  [0]{\begingroup\@sanitize@url \@url }%
\providecommand \@url [1]{\endgroup\@href {#1}{\urlprefix }}%
\providecommand \urlprefix  [0]{URL }%
\providecommand \Eprint [0]{\href }%
\providecommand \doibase [0]{http://dx.doi.org/}%
\providecommand \selectlanguage [0]{\@gobble}%
\providecommand \bibinfo  [0]{\@secondoftwo}%
\providecommand \bibfield  [0]{\@secondoftwo}%
\providecommand \translation [1]{[#1]}%
\providecommand \BibitemOpen [0]{}%
\providecommand \bibitemStop [0]{}%
\providecommand \bibitemNoStop [0]{.\EOS\space}%
\providecommand \EOS [0]{\spacefactor3000\relax}%
\providecommand \BibitemShut  [1]{\csname bibitem#1\endcsname}%
\let\auto@bib@innerbib\@empty
\bibitem [{\citenamefont {DiVincenzo}(2000)}]{DiVincenzo2000}%
  \BibitemOpen
  \bibfield  {author} {\bibinfo {author} {\bibfnamefont {D.~P.}\ \bibnamefont
  {DiVincenzo}},\ }\href {\doibase
  10.1002/1521-3978(200009)48:9/11<771::AID-PROP771>3.0.CO;2-E} {\bibfield
  {journal} {\bibinfo  {journal} {Fortschritte der Physik}\ }\textbf {\bibinfo
  {volume} {48}},\ \bibinfo {pages} {771} (\bibinfo {year} {2000})}\BibitemShut
  {NoStop}%
\bibitem [{\citenamefont {Piotrowicz}\ \emph {et~al.}(2013)\citenamefont
  {Piotrowicz}, \citenamefont {Lichtman}, \citenamefont {Maller}, \citenamefont
  {Li}, \citenamefont {Zhang}, \citenamefont {Isenhower},\ and\ \citenamefont
  {Saffman}}]{Piotrowicz2013}%
  \BibitemOpen
  \bibfield  {author} {\bibinfo {author} {\bibfnamefont {M.~J.}\ \bibnamefont
  {Piotrowicz}}, \bibinfo {author} {\bibfnamefont {M.}~\bibnamefont
  {Lichtman}}, \bibinfo {author} {\bibfnamefont {K.}~\bibnamefont {Maller}},
  \bibinfo {author} {\bibfnamefont {G.}~\bibnamefont {Li}}, \bibinfo {author}
  {\bibfnamefont {S.}~\bibnamefont {Zhang}}, \bibinfo {author} {\bibfnamefont
  {L.}~\bibnamefont {Isenhower}}, \ and\ \bibinfo {author} {\bibfnamefont
  {M.}~\bibnamefont {Saffman}},\ }\href {\doibase 10.1103/PhysRevA.88.013420}
  {\bibfield  {journal} {\bibinfo  {journal} {Phys. Rev. A}\ }\textbf {\bibinfo
  {volume} {88}},\ \bibinfo {pages} {013420} (\bibinfo {year}
  {2013})}\BibitemShut {NoStop}%
\bibitem [{\citenamefont {Xia}\ \emph {et~al.}(2015)\citenamefont {Xia},
  \citenamefont {Lichtman}, \citenamefont {Maller}, \citenamefont {Carr},
  \citenamefont {Piotrowicz}, \citenamefont {Isenhower},\ and\ \citenamefont
  {Saffman}}]{Xia2015}%
  \BibitemOpen
  \bibfield  {author} {\bibinfo {author} {\bibfnamefont {T.}~\bibnamefont
  {Xia}}, \bibinfo {author} {\bibfnamefont {M.}~\bibnamefont {Lichtman}},
  \bibinfo {author} {\bibfnamefont {K.}~\bibnamefont {Maller}}, \bibinfo
  {author} {\bibfnamefont {A.~W.}\ \bibnamefont {Carr}}, \bibinfo {author}
  {\bibfnamefont {M.~J.}\ \bibnamefont {Piotrowicz}}, \bibinfo {author}
  {\bibfnamefont {L.}~\bibnamefont {Isenhower}}, \ and\ \bibinfo {author}
  {\bibfnamefont {M.}~\bibnamefont {Saffman}},\ }\href {\doibase
  10.1103/PhysRevLett.114.100503} {\bibfield  {journal} {\bibinfo  {journal}
  {Phys. Rev. Lett.}\ }\textbf {\bibinfo {volume} {114}},\ \bibinfo {pages}
  {100503} (\bibinfo {year} {2015})}\BibitemShut {NoStop}%
\bibitem [{\citenamefont {Lukin}\ \emph {et~al.}(2001)\citenamefont {Lukin},
  \citenamefont {Fleischhauer}, \citenamefont {C\^ot\'e}, \citenamefont {Duan},
  \citenamefont {Jaksch}, \citenamefont {Cirac},\ and\ \citenamefont
  {Zoller}}]{Lukin2001}%
  \BibitemOpen
  \bibfield  {author} {\bibinfo {author} {\bibfnamefont {M.~D.}\ \bibnamefont
  {Lukin}}, \bibinfo {author} {\bibfnamefont {M.}~\bibnamefont {Fleischhauer}},
  \bibinfo {author} {\bibfnamefont {R.}~\bibnamefont {C\^ot\'e}}, \bibinfo
  {author} {\bibfnamefont {L.~M.}\ \bibnamefont {Duan}}, \bibinfo {author}
  {\bibfnamefont {D.}~\bibnamefont {Jaksch}}, \bibinfo {author} {\bibfnamefont
  {J.~I.}\ \bibnamefont {Cirac}}, \ and\ \bibinfo {author} {\bibfnamefont
  {P.}~\bibnamefont {Zoller}},\ }\href {\doibase 10.1103/PhysRevLett.87.037901}
  {\bibfield  {journal} {\bibinfo  {journal} {Phys. Rev. Lett.}\ }\textbf
  {\bibinfo {volume} {87}},\ \bibinfo {pages} {037901} (\bibinfo {year}
  {2001})}\BibitemShut {NoStop}%
\bibitem [{\citenamefont {Saffman}\ \emph {et~al.}(2010)\citenamefont
  {Saffman}, \citenamefont {Walker},\ and\ \citenamefont
  {M\o{}lmer}}]{Saffman2010}%
  \BibitemOpen
  \bibfield  {author} {\bibinfo {author} {\bibfnamefont {M.}~\bibnamefont
  {Saffman}}, \bibinfo {author} {\bibfnamefont {T.~G.}\ \bibnamefont {Walker}},
  \ and\ \bibinfo {author} {\bibfnamefont {K.}~\bibnamefont {M\o{}lmer}},\
  }\href {\doibase 10.1103/RevModPhys.82.2313} {\bibfield  {journal} {\bibinfo
  {journal} {Rev. Mod. Phys.}\ }\textbf {\bibinfo {volume} {82}},\ \bibinfo
  {pages} {2313} (\bibinfo {year} {2010})}\BibitemShut {NoStop}%
\bibitem [{\citenamefont {Stanojevic}\ and\ \citenamefont
  {C\^ot\'e}(2009)}]{Stanojevic2009}%
  \BibitemOpen
  \bibfield  {author} {\bibinfo {author} {\bibfnamefont {J.}~\bibnamefont
  {Stanojevic}}\ and\ \bibinfo {author} {\bibfnamefont {R.}~\bibnamefont
  {C\^ot\'e}},\ }\href {\doibase 10.1103/PhysRevA.80.033418} {\bibfield
  {journal} {\bibinfo  {journal} {Phys. Rev. A}\ }\textbf {\bibinfo {volume}
  {80}},\ \bibinfo {pages} {033418} (\bibinfo {year} {2009})}\BibitemShut
  {NoStop}%
\bibitem [{\citenamefont {Heidemann}\ \emph {et~al.}(2007)\citenamefont
  {Heidemann}, \citenamefont {Raitzsch}, \citenamefont {Bendkowsky},
  \citenamefont {Butscher}, \citenamefont {L\"ow}, \citenamefont {Santos},\
  and\ \citenamefont {Pfau}}]{Heidemann2007}%
  \BibitemOpen
  \bibfield  {author} {\bibinfo {author} {\bibfnamefont {R.}~\bibnamefont
  {Heidemann}}, \bibinfo {author} {\bibfnamefont {U.}~\bibnamefont {Raitzsch}},
  \bibinfo {author} {\bibfnamefont {V.}~\bibnamefont {Bendkowsky}}, \bibinfo
  {author} {\bibfnamefont {B.}~\bibnamefont {Butscher}}, \bibinfo {author}
  {\bibfnamefont {R.}~\bibnamefont {L\"ow}}, \bibinfo {author} {\bibfnamefont
  {L.}~\bibnamefont {Santos}}, \ and\ \bibinfo {author} {\bibfnamefont
  {T.}~\bibnamefont {Pfau}},\ }\href {\doibase 10.1103/PhysRevLett.99.163601}
  {\bibfield  {journal} {\bibinfo  {journal} {Phys. Rev. Lett.}\ }\textbf
  {\bibinfo {volume} {99}},\ \bibinfo {pages} {163601} (\bibinfo {year}
  {2007})}\BibitemShut {NoStop}%
\bibitem [{\citenamefont {Urban}\ \emph {et~al.}(2009)\citenamefont {Urban},
  \citenamefont {Johnson}, \citenamefont {Henage}, \citenamefont {Isenhower},
  \citenamefont {Yavuz}, \citenamefont {Walker},\ and\ \citenamefont
  {Saffman}}]{Urban2009}%
  \BibitemOpen
  \bibfield  {author} {\bibinfo {author} {\bibfnamefont {E.}~\bibnamefont
  {Urban}}, \bibinfo {author} {\bibfnamefont {T.~A.}\ \bibnamefont {Johnson}},
  \bibinfo {author} {\bibfnamefont {T.}~\bibnamefont {Henage}}, \bibinfo
  {author} {\bibfnamefont {L.}~\bibnamefont {Isenhower}}, \bibinfo {author}
  {\bibfnamefont {D.~D.}\ \bibnamefont {Yavuz}}, \bibinfo {author}
  {\bibfnamefont {T.~G.}\ \bibnamefont {Walker}}, \ and\ \bibinfo {author}
  {\bibfnamefont {M.}~\bibnamefont {Saffman}},\ }\href {\doibase DOI:
  10.1038/NPHYS1178} {\bibfield  {journal} {\bibinfo  {journal} {Nature
  Physics}\ }\textbf {\bibinfo {volume} {5}},\ \bibinfo {pages} {110} (\bibinfo
  {year} {2009})}\BibitemShut {NoStop}%
\bibitem [{\citenamefont {Wilk}\ \emph {et~al.}(2010)\citenamefont {Wilk},
  \citenamefont {Ga\"etan}, \citenamefont {Evellin}, \citenamefont {Wolters},
  \citenamefont {Miroshnychenko}, \citenamefont {Grangier},\ and\ \citenamefont
  {Browaeys}}]{Wilk2010}%
  \BibitemOpen
  \bibfield  {author} {\bibinfo {author} {\bibfnamefont {T.}~\bibnamefont
  {Wilk}}, \bibinfo {author} {\bibfnamefont {A.}~\bibnamefont {Ga\"etan}},
  \bibinfo {author} {\bibfnamefont {C.}~\bibnamefont {Evellin}}, \bibinfo
  {author} {\bibfnamefont {J.}~\bibnamefont {Wolters}}, \bibinfo {author}
  {\bibfnamefont {Y.}~\bibnamefont {Miroshnychenko}}, \bibinfo {author}
  {\bibfnamefont {P.}~\bibnamefont {Grangier}}, \ and\ \bibinfo {author}
  {\bibfnamefont {A.}~\bibnamefont {Browaeys}},\ }\href {\doibase
  10.1103/PhysRevLett.104.010502} {\bibfield  {journal} {\bibinfo  {journal}
  {Phys. Rev. Lett.}\ }\textbf {\bibinfo {volume} {104}},\ \bibinfo {pages}
  {010502} (\bibinfo {year} {2010})}\BibitemShut {NoStop}%
\bibitem [{\citenamefont {Dudin}\ and\ \citenamefont
  {Kuzmich}(2012)}]{Dudin2012}%
  \BibitemOpen
  \bibfield  {author} {\bibinfo {author} {\bibfnamefont {Y.~O.}\ \bibnamefont
  {Dudin}}\ and\ \bibinfo {author} {\bibfnamefont {A.}~\bibnamefont
  {Kuzmich}},\ }\href {\doibase DOI: 10.1038/NPHYS2413} {\bibfield  {journal}
  {\bibinfo  {journal} {Science}\ }\textbf {\bibinfo {volume} {336}},\ \bibinfo
  {pages} {887} (\bibinfo {year} {2012})}\BibitemShut {NoStop}%
\bibitem [{\citenamefont {Ebert}\ \emph {et~al.}(2014)\citenamefont {Ebert},
  \citenamefont {Gill}, \citenamefont {Gibbons}, \citenamefont {Zhang},
  \citenamefont {Saffman},\ and\ \citenamefont {Walker}}]{Ebert2014}%
  \BibitemOpen
  \bibfield  {author} {\bibinfo {author} {\bibfnamefont {M.}~\bibnamefont
  {Ebert}}, \bibinfo {author} {\bibfnamefont {A.}~\bibnamefont {Gill}},
  \bibinfo {author} {\bibfnamefont {M.}~\bibnamefont {Gibbons}}, \bibinfo
  {author} {\bibfnamefont {X.}~\bibnamefont {Zhang}}, \bibinfo {author}
  {\bibfnamefont {M.}~\bibnamefont {Saffman}}, \ and\ \bibinfo {author}
  {\bibfnamefont {T.~G.}\ \bibnamefont {Walker}},\ }\href {\doibase
  10.1103/PhysRevLett.112.043602} {\bibfield  {journal} {\bibinfo  {journal}
  {Phys. Rev. Lett.}\ }\textbf {\bibinfo {volume} {112}},\ \bibinfo {pages}
  {043602} (\bibinfo {year} {2014})}\BibitemShut {NoStop}%
\bibitem [{\citenamefont {Zeiher}\ \emph {et~al.}(2015)\citenamefont {Zeiher},
  \citenamefont {Schau\ss{}}, \citenamefont {Hild}, \citenamefont {Macr\`{\i}},
  \citenamefont {Bloch},\ and\ \citenamefont {Gross}}]{Zeiner2015}%
  \BibitemOpen
  \bibfield  {author} {\bibinfo {author} {\bibfnamefont {J.}~\bibnamefont
  {Zeiher}}, \bibinfo {author} {\bibfnamefont {P.}~\bibnamefont {Schau\ss{}}},
  \bibinfo {author} {\bibfnamefont {S.}~\bibnamefont {Hild}}, \bibinfo {author}
  {\bibfnamefont {T.}~\bibnamefont {Macr\`{\i}}}, \bibinfo {author}
  {\bibfnamefont {I.}~\bibnamefont {Bloch}}, \ and\ \bibinfo {author}
  {\bibfnamefont {C.}~\bibnamefont {Gross}},\ }\href {\doibase
  10.1103/PhysRevX.5.031015} {\bibfield  {journal} {\bibinfo  {journal} {Phys.
  Rev. X}\ }\textbf {\bibinfo {volume} {5}},\ \bibinfo {pages} {031015}
  (\bibinfo {year} {2015})}\BibitemShut {NoStop}%
\bibitem [{\citenamefont {Beterov}\ \emph {et~al.}(2011)\citenamefont
  {Beterov}, \citenamefont {Tretyakov}, \citenamefont {Entin}, \citenamefont
  {Yakshina}, \citenamefont {Ryabtsev}, \citenamefont {MacCormick},\ and\
  \citenamefont {Bergamini}}]{Beterov2011}%
  \BibitemOpen
  \bibfield  {author} {\bibinfo {author} {\bibfnamefont {I.~I.}\ \bibnamefont
  {Beterov}}, \bibinfo {author} {\bibfnamefont {D.~B.}\ \bibnamefont
  {Tretyakov}}, \bibinfo {author} {\bibfnamefont {V.~M.}\ \bibnamefont
  {Entin}}, \bibinfo {author} {\bibfnamefont {E.~A.}\ \bibnamefont {Yakshina}},
  \bibinfo {author} {\bibfnamefont {I.~I.}\ \bibnamefont {Ryabtsev}}, \bibinfo
  {author} {\bibfnamefont {C.}~\bibnamefont {MacCormick}}, \ and\ \bibinfo
  {author} {\bibfnamefont {S.}~\bibnamefont {Bergamini}},\ }\href {\doibase
  10.1103/PhysRevA.84.023413} {\bibfield  {journal} {\bibinfo  {journal} {Phys.
  Rev. A}\ }\textbf {\bibinfo {volume} {84}},\ \bibinfo {pages} {023413}
  (\bibinfo {year} {2011})}\BibitemShut {NoStop}%
\bibitem [{\citenamefont {Beterov}\ \emph {et~al.}(2013)\citenamefont
  {Beterov}, \citenamefont {Saffman}, \citenamefont {Yakshina}, \citenamefont
  {Zhukov}, \citenamefont {Tretyakov}, \citenamefont {Entin}, \citenamefont
  {Ryabtsev}, \citenamefont {Mansell}, \citenamefont {MacCormick},
  \citenamefont {Bergamini},\ and\ \citenamefont {Fedoruk}}]{Beterov2013}%
  \BibitemOpen
  \bibfield  {author} {\bibinfo {author} {\bibfnamefont {I.~I.}\ \bibnamefont
  {Beterov}}, \bibinfo {author} {\bibfnamefont {M.}~\bibnamefont {Saffman}},
  \bibinfo {author} {\bibfnamefont {E.~A.}\ \bibnamefont {Yakshina}}, \bibinfo
  {author} {\bibfnamefont {V.~P.}\ \bibnamefont {Zhukov}}, \bibinfo {author}
  {\bibfnamefont {D.~B.}\ \bibnamefont {Tretyakov}}, \bibinfo {author}
  {\bibfnamefont {V.~M.}\ \bibnamefont {Entin}}, \bibinfo {author}
  {\bibfnamefont {I.~I.}\ \bibnamefont {Ryabtsev}}, \bibinfo {author}
  {\bibfnamefont {C.~W.}\ \bibnamefont {Mansell}}, \bibinfo {author}
  {\bibfnamefont {C.}~\bibnamefont {MacCormick}}, \bibinfo {author}
  {\bibfnamefont {S.}~\bibnamefont {Bergamini}}, \ and\ \bibinfo {author}
  {\bibfnamefont {M.~P.}\ \bibnamefont {Fedoruk}},\ }\href {\doibase
  10.1103/PhysRevA.88.010303} {\bibfield  {journal} {\bibinfo  {journal} {Phys.
  Rev. A}\ }\textbf {\bibinfo {volume} {88}},\ \bibinfo {pages} {010303(R)}
  (\bibinfo {year} {2013})}\BibitemShut {NoStop}%
\bibitem [{\citenamefont {Beterov}\ \emph
  {et~al.}(2014{\natexlab{a}})\citenamefont {Beterov}, \citenamefont {Saffman},
  \citenamefont {Yakshina}, \citenamefont {Zhukov}, \citenamefont {Tretyakov},
  \citenamefont {Entin}, \citenamefont {Ryabtsev}, \citenamefont {Mansell},
  \citenamefont {MacCormick}, \citenamefont {Bergamini},\ and\ \citenamefont
  {Fedoruk}}]{Beterov2014}%
  \BibitemOpen
  \bibfield  {author} {\bibinfo {author} {\bibfnamefont {I.~I.}\ \bibnamefont
  {Beterov}}, \bibinfo {author} {\bibfnamefont {M.}~\bibnamefont {Saffman}},
  \bibinfo {author} {\bibfnamefont {E.~A.}\ \bibnamefont {Yakshina}}, \bibinfo
  {author} {\bibfnamefont {V.~P.}\ \bibnamefont {Zhukov}}, \bibinfo {author}
  {\bibfnamefont {D.~B.}\ \bibnamefont {Tretyakov}}, \bibinfo {author}
  {\bibfnamefont {V.~M.}\ \bibnamefont {Entin}}, \bibinfo {author}
  {\bibfnamefont {I.~I.}\ \bibnamefont {Ryabtsev}}, \bibinfo {author}
  {\bibfnamefont {C.~W.}\ \bibnamefont {Mansell}}, \bibinfo {author}
  {\bibfnamefont {C.}~\bibnamefont {MacCormick}}, \bibinfo {author}
  {\bibfnamefont {S.}~\bibnamefont {Bergamini}}, \ and\ \bibinfo {author}
  {\bibfnamefont {M.~P.}\ \bibnamefont {Fedoruk}},\ }\href {\doibase
  doi:10.1088/1054-660X/24/7/074013} {\bibfield  {journal} {\bibinfo  {journal}
  {Laser Physics}\ }\textbf {\bibinfo {volume} {24}},\ \bibinfo {pages}
  {074013} (\bibinfo {year} {2014}{\natexlab{a}})}\BibitemShut {NoStop}%
\bibitem [{\citenamefont {Nielsen}\ and\ \citenamefont
  {Chuang}(2011)}]{Nielsen2011}%
  \BibitemOpen
  \bibfield  {author} {\bibinfo {author} {\bibfnamefont {M.}~\bibnamefont
  {Nielsen}}\ and\ \bibinfo {author} {\bibfnamefont {I.}~\bibnamefont
  {Chuang}},\ }\href@noop {} {\emph {\bibinfo {title} {Quantum Computation and
  Quantum Information}}}\ (\bibinfo  {publisher} {Cambrige University Press},\
  \bibinfo {year} {2011})\BibitemShut {NoStop}%
\bibitem [{\citenamefont {James}\ \emph {et~al.}(2001)\citenamefont {James},
  \citenamefont {Kwiat}, \citenamefont {Munro},\ and\ \citenamefont
  {White}}]{James2001}%
  \BibitemOpen
  \bibfield  {author} {\bibinfo {author} {\bibfnamefont {D.~F.~V.}\
  \bibnamefont {James}}, \bibinfo {author} {\bibfnamefont {P.~G.}\ \bibnamefont
  {Kwiat}}, \bibinfo {author} {\bibfnamefont {W.~J.}\ \bibnamefont {Munro}}, \
  and\ \bibinfo {author} {\bibfnamefont {A.~G.}\ \bibnamefont {White}},\ }\href
  {\doibase 10.1103/PhysRevA.64.052312} {\bibfield  {journal} {\bibinfo
  {journal} {Phys. Rev. A}\ }\textbf {\bibinfo {volume} {64}},\ \bibinfo
  {pages} {052312} (\bibinfo {year} {2001})}\BibitemShut {NoStop}%
\bibitem [{\citenamefont {Poyatos}\ \emph {et~al.}(1997)\citenamefont
  {Poyatos}, \citenamefont {Cirac},\ and\ \citenamefont
  {Zoller}}]{Poyatos1997}%
  \BibitemOpen
  \bibfield  {author} {\bibinfo {author} {\bibfnamefont {J.~F.}\ \bibnamefont
  {Poyatos}}, \bibinfo {author} {\bibfnamefont {J.~I.}\ \bibnamefont {Cirac}},
  \ and\ \bibinfo {author} {\bibfnamefont {P.}~\bibnamefont {Zoller}},\ }\href
  {\doibase 10.1103/PhysRevLett.78.390} {\bibfield  {journal} {\bibinfo
  {journal} {Phys. Rev. Lett.}\ }\textbf {\bibinfo {volume} {78}},\ \bibinfo
  {pages} {390} (\bibinfo {year} {1997})}\BibitemShut {NoStop}%
\bibitem [{\citenamefont {White}\ \emph {et~al.}(2007)\citenamefont {White},
  \citenamefont {Gilchrist}, \citenamefont {Pryde}, \citenamefont {O’Brien},
  \citenamefont {Bremner},\ and\ \citenamefont {Langford}}]{White2007}%
  \BibitemOpen
  \bibfield  {author} {\bibinfo {author} {\bibfnamefont {A.~G.}\ \bibnamefont
  {White}}, \bibinfo {author} {\bibfnamefont {A.}~\bibnamefont {Gilchrist}},
  \bibinfo {author} {\bibfnamefont {G.~J.}\ \bibnamefont {Pryde}}, \bibinfo
  {author} {\bibfnamefont {J.~L.}\ \bibnamefont {O’Brien}}, \bibinfo {author}
  {\bibfnamefont {M.~J.}\ \bibnamefont {Bremner}}, \ and\ \bibinfo {author}
  {\bibfnamefont {N.~K.}\ \bibnamefont {Langford}},\ }\href {\doibase
  10.1364/JOSAB.24.000172} {\bibfield  {journal} {\bibinfo  {journal} {Journal
  of Optical Society of America B}\ }\textbf {\bibinfo {volume} {24}},\
  \bibinfo {pages} {172} (\bibinfo {year} {2007})}\BibitemShut {NoStop}%
\bibitem [{\citenamefont {Riebe}\ \emph {et~al.}(2006)\citenamefont {Riebe},
  \citenamefont {Kim}, \citenamefont {Schindler}, \citenamefont {Monz},
  \citenamefont {Schmidt}, \citenamefont {K\"orber}, \citenamefont {H\"ansel},
  \citenamefont {H\"affner}, \citenamefont {Roos},\ and\ \citenamefont
  {Blatt}}]{Riebe2006}%
  \BibitemOpen
  \bibfield  {author} {\bibinfo {author} {\bibfnamefont {M.}~\bibnamefont
  {Riebe}}, \bibinfo {author} {\bibfnamefont {K.}~\bibnamefont {Kim}}, \bibinfo
  {author} {\bibfnamefont {P.}~\bibnamefont {Schindler}}, \bibinfo {author}
  {\bibfnamefont {T.}~\bibnamefont {Monz}}, \bibinfo {author} {\bibfnamefont
  {P.~O.}\ \bibnamefont {Schmidt}}, \bibinfo {author} {\bibfnamefont {T.~K.}\
  \bibnamefont {K\"orber}}, \bibinfo {author} {\bibfnamefont {W.}~\bibnamefont
  {H\"ansel}}, \bibinfo {author} {\bibfnamefont {H.}~\bibnamefont {H\"affner}},
  \bibinfo {author} {\bibfnamefont {C.~F.}\ \bibnamefont {Roos}}, \ and\
  \bibinfo {author} {\bibfnamefont {R.}~\bibnamefont {Blatt}},\ }\href
  {\doibase 10.1103/PhysRevLett.97.220407} {\bibfield  {journal} {\bibinfo
  {journal} {Phys. Rev. Lett.}\ }\textbf {\bibinfo {volume} {97}},\ \bibinfo
  {pages} {220407} (\bibinfo {year} {2006})}\BibitemShut {NoStop}%
\bibitem [{\citenamefont {Roos}\ \emph {et~al.}(2004)\citenamefont {Roos},
  \citenamefont {Lancaster}, \citenamefont {Riebe}, \citenamefont {H\"affner},
  \citenamefont {H\"ansel}, \citenamefont {Gulde}, \citenamefont {Becher},
  \citenamefont {Eschner}, \citenamefont {Schmidt-Kaler},\ and\ \citenamefont
  {Blatt}}]{Roos2004}%
  \BibitemOpen
  \bibfield  {author} {\bibinfo {author} {\bibfnamefont {C.~F.}\ \bibnamefont
  {Roos}}, \bibinfo {author} {\bibfnamefont {G.~P.~T.}\ \bibnamefont
  {Lancaster}}, \bibinfo {author} {\bibfnamefont {M.}~\bibnamefont {Riebe}},
  \bibinfo {author} {\bibfnamefont {H.}~\bibnamefont {H\"affner}}, \bibinfo
  {author} {\bibfnamefont {W.}~\bibnamefont {H\"ansel}}, \bibinfo {author}
  {\bibfnamefont {S.}~\bibnamefont {Gulde}}, \bibinfo {author} {\bibfnamefont
  {C.}~\bibnamefont {Becher}}, \bibinfo {author} {\bibfnamefont
  {J.}~\bibnamefont {Eschner}}, \bibinfo {author} {\bibfnamefont
  {F.}~\bibnamefont {Schmidt-Kaler}}, \ and\ \bibinfo {author} {\bibfnamefont
  {R.}~\bibnamefont {Blatt}},\ }\href {\doibase 10.1103/PhysRevLett.92.220402}
  {\bibfield  {journal} {\bibinfo  {journal} {Phys. Rev. Lett.}\ }\textbf
  {\bibinfo {volume} {92}},\ \bibinfo {pages} {220402} (\bibinfo {year}
  {2004})}\BibitemShut {NoStop}%
\bibitem [{\citenamefont {Yamamoto}\ \emph {et~al.}(2010)\citenamefont
  {Yamamoto}, \citenamefont {Neeley}, \citenamefont {Lucero}, \citenamefont
  {Bialczak}, \citenamefont {Kelly}, \citenamefont {Lenander}, \citenamefont
  {Mariantoni}, \citenamefont {O'Connell}, \citenamefont {Sank}, \citenamefont
  {Wang}, \citenamefont {Weides}, \citenamefont {Wenner}, \citenamefont {Yin},
  \citenamefont {Cleland},\ and\ \citenamefont {Martinis}}]{Yamamoto2010}%
  \BibitemOpen
  \bibfield  {author} {\bibinfo {author} {\bibfnamefont {T.}~\bibnamefont
  {Yamamoto}}, \bibinfo {author} {\bibfnamefont {M.}~\bibnamefont {Neeley}},
  \bibinfo {author} {\bibfnamefont {E.}~\bibnamefont {Lucero}}, \bibinfo
  {author} {\bibfnamefont {R.~C.}\ \bibnamefont {Bialczak}}, \bibinfo {author}
  {\bibfnamefont {J.}~\bibnamefont {Kelly}}, \bibinfo {author} {\bibfnamefont
  {M.}~\bibnamefont {Lenander}}, \bibinfo {author} {\bibfnamefont
  {M.}~\bibnamefont {Mariantoni}}, \bibinfo {author} {\bibfnamefont {A.~D.}\
  \bibnamefont {O'Connell}}, \bibinfo {author} {\bibfnamefont {D.}~\bibnamefont
  {Sank}}, \bibinfo {author} {\bibfnamefont {H.}~\bibnamefont {Wang}}, \bibinfo
  {author} {\bibfnamefont {M.}~\bibnamefont {Weides}}, \bibinfo {author}
  {\bibfnamefont {J.}~\bibnamefont {Wenner}}, \bibinfo {author} {\bibfnamefont
  {Y.}~\bibnamefont {Yin}}, \bibinfo {author} {\bibfnamefont {A.~N.}\
  \bibnamefont {Cleland}}, \ and\ \bibinfo {author} {\bibfnamefont {J.~M.}\
  \bibnamefont {Martinis}},\ }\href {\doibase 10.1103/PhysRevB.82.184515}
  {\bibfield  {journal} {\bibinfo  {journal} {Phys. Rev. B}\ }\textbf {\bibinfo
  {volume} {82}},\ \bibinfo {pages} {184515} (\bibinfo {year}
  {2010})}\BibitemShut {NoStop}%
\bibitem [{\citenamefont {Shukla}\ and\ \citenamefont
  {Mahesh}(2014)}]{Shukla2014}%
  \BibitemOpen
  \bibfield  {author} {\bibinfo {author} {\bibfnamefont {A.}~\bibnamefont
  {Shukla}}\ and\ \bibinfo {author} {\bibfnamefont {T.~S.}\ \bibnamefont
  {Mahesh}},\ }\href {\doibase 10.1103/PhysRevA.90.052301} {\bibfield
  {journal} {\bibinfo  {journal} {Phys. Rev. A}\ }\textbf {\bibinfo {volume}
  {90}},\ \bibinfo {pages} {052301} (\bibinfo {year} {2014})}\BibitemShut
  {NoStop}%
\bibitem [{\citenamefont {Childs}\ \emph {et~al.}(2001)\citenamefont {Childs},
  \citenamefont {Chuang},\ and\ \citenamefont {Leung}}]{Childs2001}%
  \BibitemOpen
  \bibfield  {author} {\bibinfo {author} {\bibfnamefont {A.~M.}\ \bibnamefont
  {Childs}}, \bibinfo {author} {\bibfnamefont {I.~L.}\ \bibnamefont {Chuang}},
  \ and\ \bibinfo {author} {\bibfnamefont {D.~W.}\ \bibnamefont {Leung}},\
  }\href {\doibase 10.1103/PhysRevA.64.012314} {\bibfield  {journal} {\bibinfo
  {journal} {Phys. Rev. A}\ }\textbf {\bibinfo {volume} {64}},\ \bibinfo
  {pages} {012314} (\bibinfo {year} {2001})}\BibitemShut {NoStop}%
\bibitem [{\citenamefont {Mitchell}\ \emph {et~al.}(2003)\citenamefont
  {Mitchell}, \citenamefont {Ellenor}, \citenamefont {Schneider},\ and\
  \citenamefont {Steinberg}}]{Mitchell2003}%
  \BibitemOpen
  \bibfield  {author} {\bibinfo {author} {\bibfnamefont {M.~W.}\ \bibnamefont
  {Mitchell}}, \bibinfo {author} {\bibfnamefont {C.~W.}\ \bibnamefont
  {Ellenor}}, \bibinfo {author} {\bibfnamefont {S.}~\bibnamefont {Schneider}},
  \ and\ \bibinfo {author} {\bibfnamefont {A.~M.}\ \bibnamefont {Steinberg}},\
  }\href {\doibase 10.1103/PhysRevLett.91.120402} {\bibfield  {journal}
  {\bibinfo  {journal} {Phys. Rev. Lett.}\ }\textbf {\bibinfo {volume} {91}},\
  \bibinfo {pages} {120402} (\bibinfo {year} {2003})}\BibitemShut {NoStop}%
\bibitem [{\citenamefont {Vasilev}\ \emph {et~al.}(2009)\citenamefont
  {Vasilev}, \citenamefont {Kuhn},\ and\ \citenamefont
  {Vitanov}}]{Vasiliev2009}%
  \BibitemOpen
  \bibfield  {author} {\bibinfo {author} {\bibfnamefont {G.~S.}\ \bibnamefont
  {Vasilev}}, \bibinfo {author} {\bibfnamefont {A.}~\bibnamefont {Kuhn}}, \
  and\ \bibinfo {author} {\bibfnamefont {N.~V.}\ \bibnamefont {Vitanov}},\
  }\href {\doibase 10.1103/PhysRevA.80.013417} {\bibfield  {journal} {\bibinfo
  {journal} {Phys. Rev. A}\ }\textbf {\bibinfo {volume} {80}},\ \bibinfo
  {pages} {013417} (\bibinfo {year} {2009})}\BibitemShut {NoStop}%
\bibitem [{\citenamefont {Bergmann}\ \emph {et~al.}(1998)\citenamefont
  {Bergmann}, \citenamefont {Theuer},\ and\ \citenamefont
  {Shore}}]{Bergmann1998}%
  \BibitemOpen
  \bibfield  {author} {\bibinfo {author} {\bibfnamefont {K.}~\bibnamefont
  {Bergmann}}, \bibinfo {author} {\bibfnamefont {H.}~\bibnamefont {Theuer}}, \
  and\ \bibinfo {author} {\bibfnamefont {B.}~\bibnamefont {Shore}},\ }\href
  {\doibase 10.1103/RevModPhys.70.1003} {\bibfield  {journal} {\bibinfo
  {journal} {Review of Modern Physics}\ }\textbf {\bibinfo {volume} {70}},\
  \bibinfo {pages} {1003} (\bibinfo {year} {1998})}\BibitemShut {NoStop}%
\bibitem [{\citenamefont {Kuznetsova}(2015)}]{Kuznetsova2015}%
  \BibitemOpen
  \bibfield  {author} {\bibinfo {author} {\bibfnamefont {E.}~\bibnamefont
  {Kuznetsova}},\ }\href {\doibase 10.1088/0953-4075/48/13/135501} {\bibfield
  {journal} {\bibinfo  {journal} {J. Phys. B}\ }\textbf {\bibinfo {volume}
  {48}},\ \bibinfo {pages} {135501} (\bibinfo {year} {2015})}\BibitemShut
  {NoStop}%
\bibitem [{\citenamefont {Liu}\ and\ \citenamefont
  {Malinovskaya}(2015)}]{Liu2015}%
  \BibitemOpen
  \bibfield  {author} {\bibinfo {author} {\bibfnamefont {G.}~\bibnamefont
  {Liu}}\ and\ \bibinfo {author} {\bibfnamefont {S.~A.}\ \bibnamefont
  {Malinovskaya}},\ }\href {\doibase 10.1088/0953-4075/48/19/194001} {\bibfield
   {journal} {\bibinfo  {journal} {J. Phys. B}\ }\textbf {\bibinfo {volume}
  {48}},\ \bibinfo {pages} {194001} (\bibinfo {year} {2015})}\BibitemShut
  {NoStop}%
\bibitem [{\citenamefont {Malinovsky}\ and\ \citenamefont
  {Krause}(2001)}]{Malinovsky2001}%
  \BibitemOpen
  \bibfield  {author} {\bibinfo {author} {\bibfnamefont {V.~S.}\ \bibnamefont
  {Malinovsky}}\ and\ \bibinfo {author} {\bibfnamefont {J.~L.}\ \bibnamefont
  {Krause}},\ }\href {\doibase 10.1007/s100530170212} {\bibfield  {journal}
  {\bibinfo  {journal} {European Physical Journal D}\ }\textbf {\bibinfo
  {volume} {14}},\ \bibinfo {pages} {147} (\bibinfo {year} {2001})}\BibitemShut
  {NoStop}%
\bibitem [{\citenamefont {Isenhower}\ \emph {et~al.}(2010)\citenamefont
  {Isenhower}, \citenamefont {Urban}, \citenamefont {Zhang}, \citenamefont
  {Gill}, \citenamefont {Henage}, \citenamefont {Johnson}, \citenamefont
  {Walker},\ and\ \citenamefont {Saffman}}]{Isenhower2010}%
  \BibitemOpen
  \bibfield  {author} {\bibinfo {author} {\bibfnamefont {L.}~\bibnamefont
  {Isenhower}}, \bibinfo {author} {\bibfnamefont {E.}~\bibnamefont {Urban}},
  \bibinfo {author} {\bibfnamefont {X.~L.}\ \bibnamefont {Zhang}}, \bibinfo
  {author} {\bibfnamefont {A.~T.}\ \bibnamefont {Gill}}, \bibinfo {author}
  {\bibfnamefont {T.}~\bibnamefont {Henage}}, \bibinfo {author} {\bibfnamefont
  {T.~A.}\ \bibnamefont {Johnson}}, \bibinfo {author} {\bibfnamefont {T.~G.}\
  \bibnamefont {Walker}}, \ and\ \bibinfo {author} {\bibfnamefont
  {M.}~\bibnamefont {Saffman}},\ }\href {\doibase
  10.1103/PhysRevLett.104.010503} {\bibfield  {journal} {\bibinfo  {journal}
  {Phys. Rev. Lett.}\ }\textbf {\bibinfo {volume} {104}},\ \bibinfo {pages}
  {010503} (\bibinfo {year} {2010})}\BibitemShut {NoStop}%
\bibitem [{\citenamefont {Zhang}\ \emph
  {et~al.}(2012{\natexlab{a}})\citenamefont {Zhang}, \citenamefont {Gill},
  \citenamefont {Isenhower}, \citenamefont {Walker},\ and\ \citenamefont
  {Saffman}}]{XZhang2012}%
  \BibitemOpen
  \bibfield  {author} {\bibinfo {author} {\bibfnamefont {X.~L.}\ \bibnamefont
  {Zhang}}, \bibinfo {author} {\bibfnamefont {A.~T.}\ \bibnamefont {Gill}},
  \bibinfo {author} {\bibfnamefont {L.}~\bibnamefont {Isenhower}}, \bibinfo
  {author} {\bibfnamefont {T.~G.}\ \bibnamefont {Walker}}, \ and\ \bibinfo
  {author} {\bibfnamefont {M.}~\bibnamefont {Saffman}},\ }\href {\doibase
  10.1103/PhysRevA.85.042310} {\bibfield  {journal} {\bibinfo  {journal} {Phys.
  Rev. A}\ }\textbf {\bibinfo {volume} {85}},\ \bibinfo {pages} {042310}
  (\bibinfo {year} {2012}{\natexlab{a}})}\BibitemShut {NoStop}%
\bibitem [{\citenamefont {O'Brien}\ \emph {et~al.}(2004)\citenamefont
  {O'Brien}, \citenamefont {Pryde}, \citenamefont {Gilchrist}, \citenamefont
  {James}, \citenamefont {Langford}, \citenamefont {Ralph},\ and\ \citenamefont
  {White}}]{OBrien2004}%
  \BibitemOpen
  \bibfield  {author} {\bibinfo {author} {\bibfnamefont {J.~L.}\ \bibnamefont
  {O'Brien}}, \bibinfo {author} {\bibfnamefont {G.~J.}\ \bibnamefont {Pryde}},
  \bibinfo {author} {\bibfnamefont {A.}~\bibnamefont {Gilchrist}}, \bibinfo
  {author} {\bibfnamefont {D.~F.~V.}\ \bibnamefont {James}}, \bibinfo {author}
  {\bibfnamefont {N.~K.}\ \bibnamefont {Langford}}, \bibinfo {author}
  {\bibfnamefont {T.~C.}\ \bibnamefont {Ralph}}, \ and\ \bibinfo {author}
  {\bibfnamefont {A.~G.}\ \bibnamefont {White}},\ }\href {\doibase
  10.1103/PhysRevLett.93.080502} {\bibfield  {journal} {\bibinfo  {journal}
  {Phys. Rev. Lett.}\ }\textbf {\bibinfo {volume} {93}},\ \bibinfo {pages}
  {080502} (\bibinfo {year} {2004})}\BibitemShut {NoStop}%
\bibitem [{\citenamefont {Howard}\ \emph {et~al.}(2006)\citenamefont {Howard},
  \citenamefont {Twamley}, \citenamefont {Wittmann}, \citenamefont {Gaebel},
  \citenamefont {Jelezko},\ and\ \citenamefont {Wrachtrup}}]{Howard2006}%
  \BibitemOpen
  \bibfield  {author} {\bibinfo {author} {\bibfnamefont {M.}~\bibnamefont
  {Howard}}, \bibinfo {author} {\bibfnamefont {J.}~\bibnamefont {Twamley}},
  \bibinfo {author} {\bibfnamefont {C.}~\bibnamefont {Wittmann}}, \bibinfo
  {author} {\bibfnamefont {T.}~\bibnamefont {Gaebel}}, \bibinfo {author}
  {\bibfnamefont {F.}~\bibnamefont {Jelezko}}, \ and\ \bibinfo {author}
  {\bibfnamefont {J.}~\bibnamefont {Wrachtrup}},\ }\href {\doibase
  10.1088/1367-2630/8/3/033} {\bibfield  {journal} {\bibinfo  {journal} {New J.
  Phys.}\ }\textbf {\bibinfo {volume} {8}},\ \bibinfo {pages} {33} (\bibinfo
  {year} {2006})}\BibitemShut {NoStop}%
\bibitem [{\citenamefont {Lvovsky}\ and\ \citenamefont
  {Raymer}(2009)}]{Lvovsky2009}%
  \BibitemOpen
  \bibfield  {author} {\bibinfo {author} {\bibfnamefont {A.}~\bibnamefont
  {Lvovsky}}\ and\ \bibinfo {author} {\bibfnamefont {M.}~\bibnamefont
  {Raymer}},\ }\href {\doibase 10.1103/RevModPhys.81.299} {\bibfield  {journal}
  {\bibinfo  {journal} {Rev. Mod. Phys.}\ }\textbf {\bibinfo {volume} {81}},\
  \bibinfo {pages} {299} (\bibinfo {year} {2009})}\BibitemShut {NoStop}%
\bibitem [{\citenamefont {Beterov}\ \emph
  {et~al.}(2014{\natexlab{b}})\citenamefont {Beterov}, \citenamefont
  {Andrijauskas}, \citenamefont {Tretyakov}, \citenamefont {Entin},
  \citenamefont {Yakshina}, \citenamefont {Ryabtsev},\ and\ \citenamefont
  {Bergamini}}]{Beterov2014a}%
  \BibitemOpen
  \bibfield  {author} {\bibinfo {author} {\bibfnamefont {I.~I.}\ \bibnamefont
  {Beterov}}, \bibinfo {author} {\bibfnamefont {T.}~\bibnamefont
  {Andrijauskas}}, \bibinfo {author} {\bibfnamefont {D.~B.}\ \bibnamefont
  {Tretyakov}}, \bibinfo {author} {\bibfnamefont {V.~M.}\ \bibnamefont
  {Entin}}, \bibinfo {author} {\bibfnamefont {E.~A.}\ \bibnamefont {Yakshina}},
  \bibinfo {author} {\bibfnamefont {I.~I.}\ \bibnamefont {Ryabtsev}}, \ and\
  \bibinfo {author} {\bibfnamefont {S.}~\bibnamefont {Bergamini}},\ }\href
  {\doibase 10.1103/PhysRevA.90.043413} {\bibfield  {journal} {\bibinfo
  {journal} {Phys. Rev. A}\ }\textbf {\bibinfo {volume} {90}},\ \bibinfo
  {pages} {043413} (\bibinfo {year} {2014}{\natexlab{b}})}\BibitemShut
  {NoStop}%
\bibitem [{\citenamefont {Ebert}\ \emph {et~al.}(2015)\citenamefont {Ebert},
  \citenamefont {Kwon}, \citenamefont {Walker},\ and\ \citenamefont
  {Saffman}}]{Ebert2015}%
  \BibitemOpen
  \bibfield  {author} {\bibinfo {author} {\bibfnamefont {M.}~\bibnamefont
  {Ebert}}, \bibinfo {author} {\bibfnamefont {M.}~\bibnamefont {Kwon}},
  \bibinfo {author} {\bibfnamefont {T.~G.}\ \bibnamefont {Walker}}, \ and\
  \bibinfo {author} {\bibfnamefont {M.}~\bibnamefont {Saffman}},\ }\href
  {\doibase 10.1103/PhysRevLett.115.093601} {\bibfield  {journal} {\bibinfo
  {journal} {Phys. Rev. Lett.}\ }\textbf {\bibinfo {volume} {115}},\ \bibinfo
  {pages} {093601} (\bibinfo {year} {2015})}\BibitemShut {NoStop}%
\bibitem [{\citenamefont {Derevianko}\ \emph {et~al.}(2015)\citenamefont
  {Derevianko}, \citenamefont {Komar}, \citenamefont {Topcu}, \citenamefont
  {Kroeze},\ and\ \citenamefont {Lukin}}]{Derevianko2015}%
  \BibitemOpen
  \bibfield  {author} {\bibinfo {author} {\bibfnamefont {A.}~\bibnamefont
  {Derevianko}}, \bibinfo {author} {\bibfnamefont {P.}~\bibnamefont {Komar}},
  \bibinfo {author} {\bibfnamefont {T.}~\bibnamefont {Topcu}}, \bibinfo
  {author} {\bibfnamefont {R.~M.}\ \bibnamefont {Kroeze}}, \ and\ \bibinfo
  {author} {\bibfnamefont {M.~D.}\ \bibnamefont {Lukin}},\ }\href {\doibase
  10.1103/PhysRevA.92.063419} {\bibfield  {journal} {\bibinfo  {journal} {Phys.
  Rev. A}\ }\textbf {\bibinfo {volume} {92}},\ \bibinfo {pages} {063419}
  (\bibinfo {year} {2015})}\BibitemShut {NoStop}%
\bibitem [{\citenamefont {Beterov}\ \emph {et~al.}(2009)\citenamefont
  {Beterov}, \citenamefont {Ryabtsev}, \citenamefont {Tretyakov},\ and\
  \citenamefont {Entin}}]{Beterov2009}%
  \BibitemOpen
  \bibfield  {author} {\bibinfo {author} {\bibfnamefont {I.~I.}\ \bibnamefont
  {Beterov}}, \bibinfo {author} {\bibfnamefont {I.~I.}\ \bibnamefont
  {Ryabtsev}}, \bibinfo {author} {\bibfnamefont {D.~B.}\ \bibnamefont
  {Tretyakov}}, \ and\ \bibinfo {author} {\bibfnamefont {V.~M.}\ \bibnamefont
  {Entin}},\ }\href {\doibase 10.1103/PhysRevA.79.052504} {\bibfield  {journal}
  {\bibinfo  {journal} {Phys. Rev. A}\ }\textbf {\bibinfo {volume} {79}},\
  \bibinfo {pages} {052504} (\bibinfo {year} {2009})}\BibitemShut {NoStop}%
\bibitem [{\citenamefont {Saffman}\ \emph {et~al.}(2011)\citenamefont
  {Saffman}, \citenamefont {Zhang}, \citenamefont {Gill}, \citenamefont
  {Isenhower},\ and\ \citenamefont {Walker}}]{Saffman2011}%
  \BibitemOpen
  \bibfield  {author} {\bibinfo {author} {\bibfnamefont {M.}~\bibnamefont
  {Saffman}}, \bibinfo {author} {\bibfnamefont {X.~L.}\ \bibnamefont {Zhang}},
  \bibinfo {author} {\bibfnamefont {A.}~\bibnamefont {Gill}}, \bibinfo {author}
  {\bibfnamefont {L.}~\bibnamefont {Isenhower}}, \ and\ \bibinfo {author}
  {\bibfnamefont {T.~G.}\ \bibnamefont {Walker}},\ }\href {\doibase
  10.1088/1742-6596/264/1/012023} {\bibfield  {journal} {\bibinfo  {journal}
  {Journal of Physics: Conference Series}\ }\textbf {\bibinfo {volume} {264}},\
  \bibinfo {pages} {012023} (\bibinfo {year} {2011})}\BibitemShut {NoStop}%
\bibitem [{\citenamefont {Honer}\ \emph {et~al.}(2011)\citenamefont {Honer},
  \citenamefont {L\"ow}, \citenamefont {Weimer}, \citenamefont {Pfau},\ and\
  \citenamefont {B\"uchler}}]{Honer2011}%
  \BibitemOpen
  \bibfield  {author} {\bibinfo {author} {\bibfnamefont {J.}~\bibnamefont
  {Honer}}, \bibinfo {author} {\bibfnamefont {R.}~\bibnamefont {L\"ow}},
  \bibinfo {author} {\bibfnamefont {H.}~\bibnamefont {Weimer}}, \bibinfo
  {author} {\bibfnamefont {T.}~\bibnamefont {Pfau}}, \ and\ \bibinfo {author}
  {\bibfnamefont {H.~P.}\ \bibnamefont {B\"uchler}},\ }\href {\doibase
  10.1103/PhysRevLett.107.093601} {\bibfield  {journal} {\bibinfo  {journal}
  {Phys. Rev. Lett.}\ }\textbf {\bibinfo {volume} {107}},\ \bibinfo {pages}
  {093601} (\bibinfo {year} {2011})}\BibitemShut {NoStop}%
\bibitem [{\citenamefont {Zhang}\ \emph
  {et~al.}(2012{\natexlab{b}})\citenamefont {Zhang}, \citenamefont {McConnell},
  \citenamefont {Cuk}, \citenamefont {Lin}, \citenamefont {Schleier-Smith},
  \citenamefont {Leroux},\ and\ \citenamefont {Vuletic}}]{HZhang2012}%
  \BibitemOpen
  \bibfield  {author} {\bibinfo {author} {\bibfnamefont {H.}~\bibnamefont
  {Zhang}}, \bibinfo {author} {\bibfnamefont {R.}~\bibnamefont {McConnell}},
  \bibinfo {author} {\bibfnamefont {S.}~\bibnamefont {Cuk}}, \bibinfo {author}
  {\bibfnamefont {Q.}~\bibnamefont {Lin}}, \bibinfo {author} {\bibfnamefont
  {M.~H.}\ \bibnamefont {Schleier-Smith}}, \bibinfo {author} {\bibfnamefont
  {I.}~\bibnamefont {Leroux}}, \ and\ \bibinfo {author} {\bibfnamefont
  {V.}~\bibnamefont {Vuletic}},\ }\href {\doibase
  10.1103/PhysRevLett.109.133603} {\bibfield  {journal} {\bibinfo  {journal}
  {Phys. Rev. Lett.}\ }\textbf {\bibinfo {volume} {109}},\ \bibinfo {pages}
  {133603} (\bibinfo {year} {2012}{\natexlab{b}})}\BibitemShut {NoStop}%
\bibitem [{\citenamefont {Tretyakov}\ \emph {et~al.}(2014)\citenamefont
  {Tretyakov}, \citenamefont {Entin}, \citenamefont {Yakshina}, \citenamefont
  {Beterov}, \citenamefont {Andreeva},\ and\ \citenamefont
  {Ryabtsev}}]{Tretyakov2014}%
  \BibitemOpen
  \bibfield  {author} {\bibinfo {author} {\bibfnamefont {D.~B.}\ \bibnamefont
  {Tretyakov}}, \bibinfo {author} {\bibfnamefont {V.~M.}\ \bibnamefont
  {Entin}}, \bibinfo {author} {\bibfnamefont {E.~A.}\ \bibnamefont {Yakshina}},
  \bibinfo {author} {\bibfnamefont {I.~I.}\ \bibnamefont {Beterov}}, \bibinfo
  {author} {\bibfnamefont {C.}~\bibnamefont {Andreeva}}, \ and\ \bibinfo
  {author} {\bibfnamefont {I.~I.}\ \bibnamefont {Ryabtsev}},\ }\href {\doibase
  10.1103/PhysRevA.90.041403} {\bibfield  {journal} {\bibinfo  {journal} {Phys.
  Rev. A}\ }\textbf {\bibinfo {volume} {90}},\ \bibinfo {pages} {041403}
  (\bibinfo {year} {2014})}\BibitemShut {NoStop}%
\bibitem [{\citenamefont {Petrosyan}\ and\ \citenamefont
  {M\o{}lmer}(2013)}]{Petrosyan2013}%
  \BibitemOpen
  \bibfield  {author} {\bibinfo {author} {\bibfnamefont {D.}~\bibnamefont
  {Petrosyan}}\ and\ \bibinfo {author} {\bibfnamefont {K.}~\bibnamefont
  {M\o{}lmer}},\ }\href {\doibase 10.1103/PhysRevA.87.033416} {\bibfield
  {journal} {\bibinfo  {journal} {Phys. Rev. A}\ }\textbf {\bibinfo {volume}
  {87}},\ \bibinfo {pages} {033416} (\bibinfo {year} {2013})}\BibitemShut
  {NoStop}%
\end{thebibliography}
\end{document}